\documentclass[preprint2]{aastex}
\usepackage{natbib}
\usepackage{graphicx}
\usepackage{color}

\shorttitle{Physical And Chemical Conditions in Centuarus A}
\shortauthors{McCoy et al.}

\begin{document}

\title{ALMA Observations of the Physical and Chemical Conditions in Centaurus A}

\author{Mark McCoy\altaffilmark{1}, J{\"u}rgen Ott\altaffilmark{2}, David S. Meier\altaffilmark{1,2}, S{\'e}bastien Muller\altaffilmark{3}, Daniel Espada\altaffilmark{4,5,6}, Sergio Mart\'in\altaffilmark{7,4}, Frank P. Israel\altaffilmark{8}, Christian Henkel\altaffilmark{9,10},  Violette Impellizzeri\altaffilmark{11,4}, Susanne Aalto\altaffilmark{3}, Philip G. Edwards\altaffilmark{12}, Andreas Brunthaler\altaffilmark{9}, Nadine Neumayer\altaffilmark{13}, Alison B. Peck\altaffilmark{11}, Paul van der Werf\altaffilmark{8}, Ilana Feain\altaffilmark{14}}


\altaffiltext{1}{New Mexico Institute of Mining and Technology, 801 Leroy Place, Socorro, NM, 87801, USA; Mark.McCoy@student.nmt.edu}
\altaffiltext{2}{National Radio Astronomy Observatory, Pete V. Domenici Array Science Center, P.O. Box O, Socorro, NM, 87801, USA}
\altaffiltext{3}{Department of Earth, Space and Environment, Chalmers University of Technology, Onsala Space Observatory, SE-43992 Onsala, Sweden}
\altaffiltext{4}{Joint Alma Observatory, Alonso de C\'ordova, 3107, Vitacura, Santiago 763-0355, Chile}
\altaffiltext{5}{National Astronomical Observatory of Japan, 2-21-1 Osawa, Mitaka, Tokyo 181-8588, Japan}
\altaffiltext{6}{The Graduate University for Advanced Studies (SOKENDAI), 2-21-1 Osawa, Mitaka, Tokyo 181-0015, Japan}
\altaffiltext{7}{European Southern Observatory, Alonso de C\'ordova 3107, Vitacura, Santiago, Chile}
\altaffiltext{8}{Leiden Observatory, Leiden University, PO Box 9513, NL-2300 RA Leiden, The Netherlands}
\altaffiltext{9}{Max-Planck-Institut f\"{u}r Radioastronomie, Auf dem H\"{u}gel 69, D-53121 Bonn, Germany}
\altaffiltext{10}{Astron. Dept., King Abdulaziz University, PO Box 80203, Jeddah 21589, Saudia Arabia}
\altaffiltext{11}{National Radio Astronomy Observatory, 520 Edgemont Rd, Charlottesville, VA 22903, USA}
\altaffiltext{12}{CSIRO CASS/ATNF, PO Box 76 Epping, NSW 1710, Australia}
\altaffiltext{13}{Max-Planck Institut f\"{u}r Astronomie, K\"{o}nigstuhl 17, D-69117 Heidelberg, Germany}
\altaffiltext{14}{School of Physics, University of Sydney, NSW 2006, Australia}

\begin{abstract}
Centaurus A, with its gas-rich elliptical host galaxy, NGC 5128, is the nearest radio galaxy at a distance of 3.8 Mpc.  Its proximity allows us to study the interaction between an active galactic nucleus, radio jets, and molecular gas in great detail. We present ALMA observations of  low J  transitions of   three  CO isotopologues, HCN, HCO$^{+}$, HNC, CN, and CCH toward the inner projected 500 pc of NGC 5128.  Our observations resolve physical sizes down to 40 pc.  By observing multiple chemical probes, we determine the physical and chemical conditions of the nuclear interstellar medium of NGC 5128.  This region contains molecular arms associated with the dust lanes and a circumnuclear disk (CND) interior to the molecular arms. The  CND  is approximately 400 pc by 200 pc and appears to be chemically distinct from the molecular arms.  It is dominated by dense gas tracers while the molecular arms are dominated by $^{12}$CO and its  rare  isotopologues. The CND has a higher temperature, elevated CN/HCN and HCN/HNC  intensity  ratios, and much weaker  $^{13}$CO and C$^{18}$O emission  than the molecular arms.  This suggests an influence from the AGN on the CND molecular gas.  There is also absorption against the AGN with a low velocity complex near the systemic velocity and a high velocity complex shifted by about 60 km s$^{-1}$.  We find  similar chemical properties between the CND in emission and both the low and high velocity absorption complexes  implying that both likely  originate from  the CND. If the HV complex does originate in the CND, then that gas  would correspond to  gas falling toward the supermassive black hole.  

\end{abstract}


\section{Introduction}

Radio galaxies are the largest contribution to the extragalactic radio sky above 1 mJy \citep{Con1984}.  Radio galaxies are fairly rare in the local universe.   Most radio galaxies are of the elliptical type \citep[e.g.][]{Wil1995} and frequently show dust features \citep[e.g.][]{Ver1999}.  In the nuclear regions of some radio galaxies, CO rotational transitions are observed with double-horned profiles indicative of rotating disks (e.g. \citealt{Leo2003,Lim2003}).  Mergers with small gas-rich galaxies may be the source of cold gas in evolved elliptical galaxies (e.g. \citealt{Sal2003,Dav2011,OSu2015}).  

The power source of the active galactic nuclei (AGNs) and the giant synchrotron jets/lobes  in radio galaxies is thought to be the accretion of material onto the central supermassive black hole.  The mechanisms for transporting gas from the outer kiloparsecs to the central  parsecs  of radio galaxies are, however, not well understood.  In turn, the AGN and jets also impact their environment, possibly quenching or inducing  star formation activity, and launching material  out of the galaxy  (e.g. \citealt{Ost2010,Aal2008,Mei2007,Ott2013}).  

High spatial resolution observations are required to determine the nature and properties of the gas in the nuclear region of radio galaxies. NGC 5128 is the nearest radio galaxy (3.8 Mpc; \citealt{Har2010}).  Its proximity makes NGC 5128 an important source for further understanding the details of accretion onto SMBHs. 

NGC 5128 is the host galaxy of the strong radio source Centaurus A (hereafter Cen A; see \citealt{Isr1998} for a comprehensive review). The $5.5 \times 10^{7}$ M$_{\odot}$ supermassive black hole at the center of NGC 5128 \citep{Cap2009,Neu2010} is powering synchrotron jets, which create the inner and outer radio lobe structures of Cen A.  The inner lobes extend roughly 12 arcminutes on the sky and are much brighter than the outer lobes \citep{Fea2011}, which extend 8 degrees on the sky.  The synchrotron jet in Cen A has been observed to interact with the gas and to induce star formation (SF) (e.g. \citealt{Cha2000,Aul2012,Cro2012,San2015,Sal2016}). 

The history of NGC 5128 is dominated by multiple merger events which  have  left signatures  in the form of a system of optical shells (e.g. \citealt{Mal1983,Pen2002}).  The galaxy is also bisected by a prominent dust lane.   Presumably, the merger events which produced the optical shells  are  also responsible for the dust lane through multiple warps and gas re-accretion.  Modeling of warps in the gas, from 6.5 kpc down to the inner 2 pc, have been compiled by \citet{Qui2006,Qui2010} .   The  warped disk model has many kinks that give rise to the various  morphological and kinematic  structures observed in Cen A, most prominently, a projected parallelogram structure detected in the IR that is associated with the molecular arms (detected in H$\alpha$: \citealt{Nic1992}; NIR: \citealt{Qui1993,Kai2009}; Sub mm continuum: \citealt{Haw1993,Lee2002}; CO: \citealt{Phi1987,Eck1990,Ryd1993,Lis2001,Esp2009}; Mid-IR: \citealt{Mir1999,Qui2006}).  H$\alpha$ emission detected in this putative warped disk indicates  that massive stars are currently forming \citep[e.g.][]{Nic1992}. A recent study, however, suggests that the gas associated with the dust parallelogram may  also  be described  by  spiral arms  in addition to  a warped disk \citep{Esp2012}.   Interior to the parallelogram structure is a 405 pc by 215 pc circumnuclear disk with a different position angle and inclination (\citealt{Isr1990,Isr1991,Isr1992,Ryd1993,Neu2007,Qui2010}).

\citet{Esp2017} describe additional molecular components  on much smaller scales within the circumnuclear disk,  including CND filaments, a nuclear ring, nuclear filaments (probably shocks), a nuclear disk ($\sim$ 30 pc in size) and absorption against the AGN. The nuclear disk is detected in ionized and molecular  hydrogen \citealt{Mar2001,Mar2006,Neu2007}).   The absorption profile consists of a low velocity complex near the systemic velocity and a high velocity complex shifted by $\sim 60$ km s$^{-1}$ (\citealt{Wik1997,Mul2009}). The location of the gas that gives rise to these absorption features is still debated. The absorption has been detected in HI (e.g. \citealt{Rob1970,Hul1983,Sar2002,Esp2011}) and molecular lines (e.g. \citealt{Isr1990,Wik1997,Eck1999,Esp2011}). 

This paper focuses on the kinematics, dynamics, and chemistry of the disk-like structure associated with the dust lane [referred to as the molecular arm feature] and the circumnuclear disk  (CND).  The sensitivity and resolution of the Atacama Large Millimeter/Submillimeter Array (ALMA) allow us to test the physical and chemical conditions in such an extreme environment.  We present a multi-line survey to derive the nuclear gas conditions and to study the impact of the AGN on the nuclear gas.    Section 2 describes the ALMA observations and the data reduction.  Section 3 presents results from the observations.  Section 4 focuses on describing the dynamics, masses and chemistry of the molecular gas, a comparison of the transitions seen in both emission and absorption, and possible signatures AGN  influence on these spectra.  Section 5 summarizes our results. 

\section{Observations and Data Reduction}

\subsection{Observational Setup}

Observations were taken with ALMA during Cycle 0 early science observing under the project code 2011.0.00010.S (PI:Ott).  We observed 20 spectral windows, four in the 1 mm band (ALMA Band 6) and sixteen in the 3 mm band (ALMA Band 3).  The observation and image parameters are outlined in Table \ref{tab:obs}. 

The 1 mm observations were  carried out  in a single run with 17 antennas in  a  compact configuration  on 2012 Jan 24,  with baselines of 18 m to 125 m.  The maximum recoverable scale is 10$"$  and the primary beam  is 27$"$.  Titan  is  the flux and bandpass calibrator and complex gain (phase/gain) calibration was done by observing PKS B1424$-$418 (J1427$-$4206) every 10 minutes.  There were no significant spectral lines in the atmosphere of Titan that may contaminate the Cen A spectra.  The total on-source integration time  is  $\sim$ 33 minutes with each spectral window separated into 3840 channels with  spacings  of 244.14 kHz per channel corresponding to a velocity  resolution of 0.66 km s$^{-1}$ after Hanning smoothing  and a bandwidth of 937.5 MHz (1260 km s$^{-1}$).

The 3 mm band observations were taken in multiple runs spanning the time between 2012 Apr 7 and 2012 May 9 in  a more  extended configuration with baselines of 36 m to 400m.  The maximum recoverable scale was 20$"$  and the primary beam was  62$"$. The 3 mm observations were performed with 17 antennas for the first two observing runs and 16 antennas for the final two observations (See Table \ref{tab:obs} for details).  As  with  the Band 6 observations, the flux and complex gain calibrators are  Titan and PKS B1424$-$418 (J1427-4206), respectively.  The phase calibrator  was  observed every 16 minutes.  3C279 (J1256$-$0547) was used as  the  bandpass calibrator. The total on-source integration time  is  $\sim$ 66 minutes with each spectral window again divided into 3840 channels but with  spacings  of 61.04 kHz per channel corresponding to a  velocity resolution between 0.32 and 0.44 km s$^{-1}$  with a bandwidth of 234.4 MHz (approximately 730 km s$^{-1}$). 

The  configurations were selected such that the  maximum recoverable scales are  approximately the same between the 3 mm and 1 mm bands,  so comparisons between the fluxes in each band should, in principle,  be robust  on the sampled scales.

\subsection{Calibration and Imaging }

Initial calibration and editing of the data were performed by the ALMA team using the CASA (Common Astronomy Software Applications) data reduction package \citep{CASAref}.  After delivery, self calibration and imaging were performed by us with CASA version 4.1.  We complemented our data with the Science Verification $^{12}$CO($2-1$) observations of Cen A, giving us a total of nine molecular transitions with emission detected in the nuclear region.  

We performed  a self calibration on the bright radio continuum core of Cen A using the CASA task \textsc{gaincal} to generate the calibration tables and \textsc{applycal} to apply the corrections to the data.  The calibration tables were made by carefully avoiding channels  with line emission. As the first step, we self-calibrated the phases only, then we did an iteration of  simultaneous amplitude and phase self-calibration.  We then used the CASA task \textsc{uvcontsub2} to fit a 1st order polynomial continuum model to channels that were free of both  line  emission and absorption.  This model was then subtracted from the  (u,v)  data to create  a  continuum-free spectral line data set for each  line.     

With the continuum emission removed we then proceeded to image each of the spectral windows.  Imaging was performed using the CASA task \textsc{clean} with natural weighting and an image size of 320 $\times$ 320 pixels at 0.2 arcseconds per pixel.  This corresponds to an image size of 64$"$ which encompasses the primary beams of both the Band 3 (62$"$ primary beam) and the Band 6 (27$"$ primary beam) observations.  The synthesized beams for the images are listed in Table \ref{tab:obs}.  Continuum images were made using the line-free channels of the two representative frequency ranges, centered on 220.007 GHz and 90.498 GHz, with synthesized beams listed in Table \ref{tab:obs}.  An overall flux accuracy of 5\% was assumed for the Band 3 observations and a flux accuracy of 10\% was assumed for the Band 6 observations\footnote{From ALMA cycle 0 capabilites. \url{https://almascience.eso.org/documents-and-tools/cycle-0/capabilities/at\_download/file$\&$usg=AOvVaw3gee33Ts0yieHWve2ubbZs}} 

Each frequency range was imaged at the full spectral resolution (3840 channels) to obtain the highest spectral resolution absorption spectra toward the continuum source of Cen A achievable with this data.  The image cubes used to make the absorption spectra were not cleaned because of an artifact created  during the execution of  the \textsc{clean} algorithm.  Near  the location of the absorption,  \textsc{clean} created a positive spike in the image cube as the map transitioned from noise to a deep absorption feature.  The clean artifacts are not fully understood, but they appear to be related to the high dynamic range change between adjacent absorption / non-absorption channels. The artifacts are not present in the dirty cube that we used for the analysis of the absorption spectra.  

Broader spectral resolution (averaged over 10 channels) image cubes were created and cleaned to map the emission.  The emission cubes were then inspected for  artifacts  created by \textsc{clean}.  The artifacts in the emission were all located in the center of the field of view, co-spatial with the absorption, which were already masked due to the absorption hole (see below) and so do not affect the displayed emission maps.  Therefore, in summary, the artifact does not impact any map / result presented in this paper.  The emission cubes were all then convolved to a common  beam of 2.25$"$ $\times$ 2.25$"$, which is slightly larger than the original synthesized beam sizes given in Table \ref{tab:obs}. The lone exception is the  $^{12}$CO($2-1$) image cube,  which due to u--v coverage was unable to be convolved to the same beam.  In order to compare the $^{12}$CO($2-1$) and the other isotopologues, the $^{12}$CO($2-1$) was convolved to a 3.0$"$ $\times$ 3.0$"$ beam and compared to separate $^{13}$CO($2-1$), $^{13}$CO($1-0$) and C$^{18}$O($2-1$) images also convolved to 3.0$"$ $\times$ 3.0$"$.  Integrated intensity,  intensity-weighted velocity and intensity-weighted  velocity dispersion maps were made  for  each transition with observed emission. A mask was created for each molecular transition by convolving the image cube with a 4.5$"$ $\times$ 4.5$"$ beam (twice the size of the synthesized beam of the image cube) and then only including emission in the original image cube where the smoothed cube contained emission stronger than $\sim$ 3 times the RMS.  This clipping was necessary to minimize the effect of the absorption near the radio core  on the rest of the CND emission.  This mask was applied to the image cube during the  map making process. In addition, a mask of $\sim 5" \times 5"$ was used to mask the absorption at the center for each of the moment maps.  

Additional ALMA data obtained during science verification (2011.0.00008.SV)  were used to provide the $^{12}$CO($2-1$) 1.3 mm transition for comparison.  The data were edited, calibrated, and imaged  in a consistent manner as the rest of our data for direct comparison.  However,  the  spectral resolution  is coarser than our data, at 10 km s$^{-1}$ per channel, while our data were imaged with velocity resolutions ranging from 1.6 to 3.3 km s$^{-1}$ per channel. 

\section{Results}

\subsection{Millimeter Continuum}
\begin{figure*}
\centering
\includegraphics[width=0.8\textwidth]{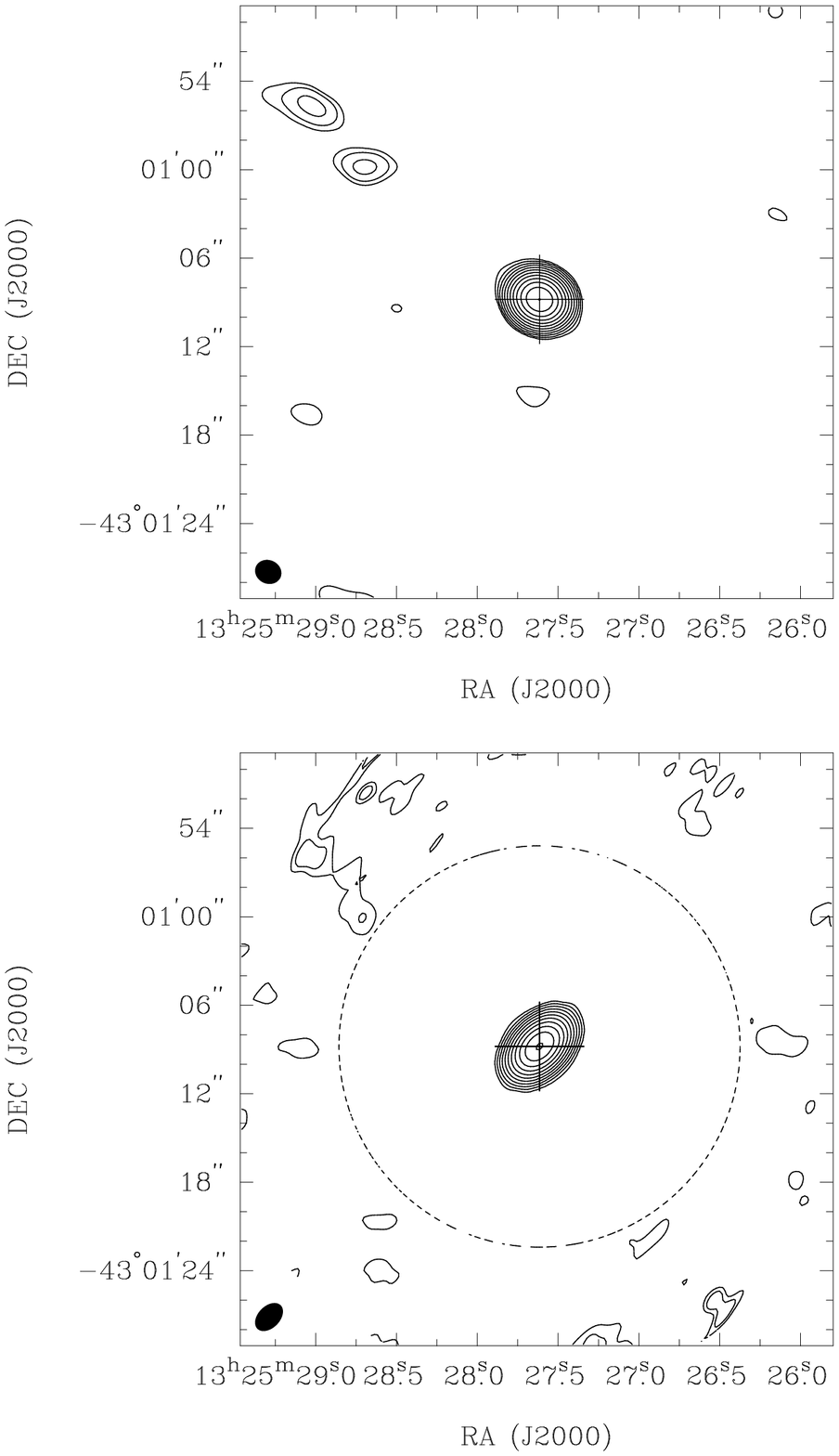}
\caption{The continuum image for 3 mm (top) and 1.3 mm (bottom).  The contours for both images are in steps of $2^{n} \cdot 3 \sigma$ with n=0,1,2,\ldots,11 and $\sigma=1.3$ mJy beam$^{-1}$  and $2.1$ mJy beam$^{-1}$  for 3 mm and 1 mm respectively.  Both images show the continuum from the central core as well as 2 knots of the northern jet.  The dashed circle in the 1 mm image shows the 50\% level of the primary beam.  The synthesized beams are shown as filled ellipses in the bottom left corner of each plot, 1.66$"$ $\times$ 1.61$"$, +59.0$^{\circ}$ for 3 mm and  2.23$"$ $\times$ 1.46$"$, $-$44.6$^{\circ}$ for 1.3 mm. The cross is 6 arcseconds in size  ($\sim$ 110 pc) and is centered on the core continuum emission  (located at RA: 13$^{h}$25$^{m}$27.616$^{s}$	DEC: -43$^{\circ}$01$'$08.813$''$ in J2000). }
\label{Fig:Cont}
\end{figure*}

 The AGN is detected as an unresolved millimeter continuum source  at a position of RA(J2000): 13$^{h}$25$^{m}$27.616$^{s}$ DEC (J2000): -43$^{\circ}$01$'$08.813$''$.   The continuum fluxes for the bright central source  were determined from apertures in cleaned continuum images and  are listed in column (7) of Table \ref{tab:obs}.  Figure \ref{Fig:Cont} shows the 3 mm band (at 90.5 GHz) and 1 mm band (at 220.0 GHz) continuum images.  The dashed line in Figure \ref{Fig:Cont}  (lower panel) shows the half power of the 1.3 mm primary beam, while the corresponding half power for 3 mm is outside the displayed field of view.  The radio jet is also detected in both bands, but it is stronger in the 3 mm continuum image.  Two knots of the northern jet are detected, but no portion of the southern jet is  seen.  This is consistent with previous observations.  The knots correspond to A1 (inner) and A2 (outer) of \citet{Cla1992}, which were studied in more detail at radio wavelengths and X-ray energies by \citet{Har2003}. Self calibration was required to achieve the dynamic range necessary to detect the jet in the presence of the strong nuclear source.  We  achieved  rms values for the 3 mm and 1 mm images of 1.3 mJy beam$^{-1}$ and 2.1 mJy beam$^{-1}$ respectively.  The jet integrated flux is around  $30 \pm 5$ mJy in band 3 and $ 20 \pm 5$ mJy in band 6 compared to the core flux of $8.1 \pm 0.4$ Jy in band 3 and $6.7 \pm 0.7$ Jy in band 6.    Both knots of the jet are well detected in both images, with the outer knot slightly stronger than the inner knot in both bands.  The spectral index between the two bands is  $-0.2 \pm 0.1$ for the bright radio core continuum, and $-0.4 \pm 0.2$ for the jet continuum,  with the spectral index $\alpha$ defined as $S \propto \nu^{\alpha}$.  \citet{Isr2008}  determine the core continuum spectral index to vary over the range -0.2 -- -0.6, consistent with our value.  However, the 3 mm and 1 mm continuum were not observed simultaneously.  This adds uncertainty to all flux comparisons due to the known $\sim$30\% variability of Centaurus A (See \citealt{Isr2008}, their Figure 5). This time variability in the flux could potentially explain the flux differences listed in Table \ref{tab:obs}. However the fluxes determined for different frequencies in the same observing band give a positive spectral index, so there are likely systemic errors as well. 

\subsection{Molecular Gas  Structure}

We detected eight molecular transitions in emission toward the nuclear region of Cen A. The detected transitions are: $J = 2-1$ rotational transitions of $^{13}$CO and C$^{18}$O [Band 6], the $J = 1-0$ rotational transitions of  $^{13}$CO, HCO$^{+}$, HCN, HNC, and the $N=1-0$; $J=\frac{3}{2}-\frac{1}{2}$ transitions of CN and CCH (hereafter CN($1-0$) and CCH($1-0$)) [Band 3].  In Band 6, HNCO$(10_{0,10}-9_{0,9})$ and H$_{2}$CO$(3_{0,3}-2_{0,2})$  were targeted but not  detected in emission. HC$_{3}$N$(10-9)$, CH$_{3}$CN$(5_{3}-4_{3})$, HNCO$(4_{0,4}-3_{0,3})$, and $J = 1-0$ of C$^{18}$O, C$^{17}$O, and N$_{2}$H$^{+}$ were targeted in band 3  (3mm), but not detected in emission.  

\begin{figure*}
\centering
\includegraphics[width=\textwidth]{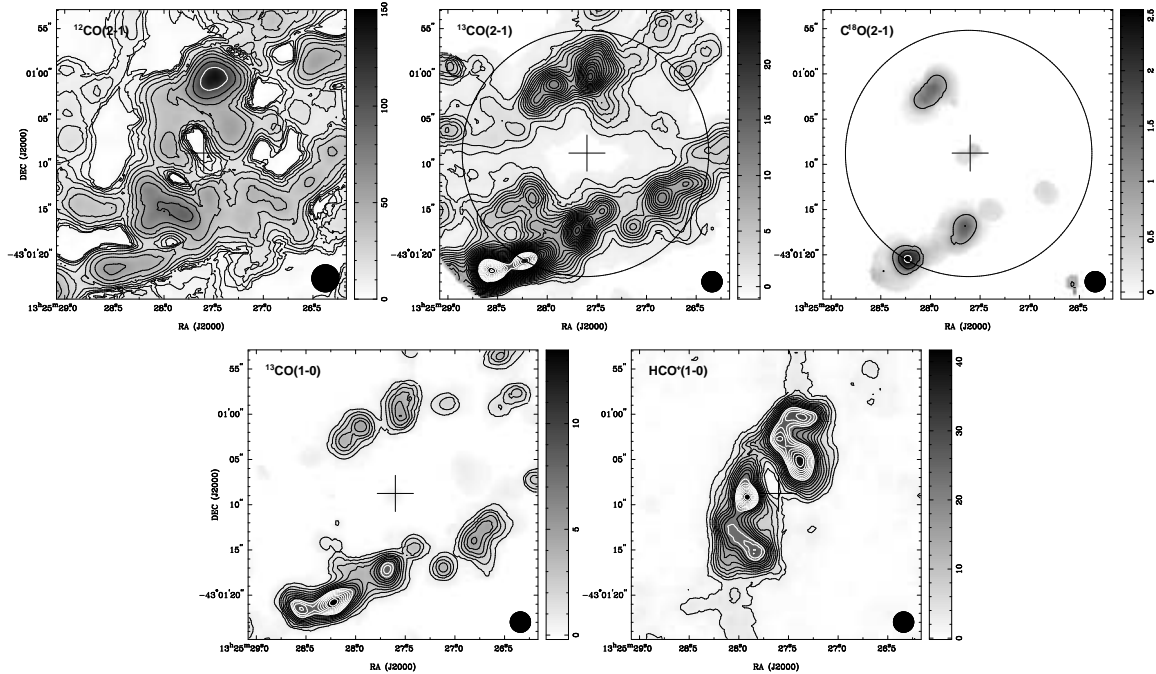}
\caption{The integrated intensity maps of CO and its isotopologues are shown in this figure along with HCO$^{+}$(1-0).   The top left panel  presents  $^{12}$CO($2-1$),  the  top middle panel   presents  $^{13}$CO($2-1$),  the  top right panel   presents C$^{18}$O($2-1$), the  bottom left panel   presents $^{13}$CO($1-0$)  and  the  bottom right panel   presents  HCO$^{+}$($1-0$).  The central cross in each of the images identifies the location of the central black hole  (located at RA: 13$^{h}$25$^{m}$27.616$^{s}$	DEC: -43$^{\circ}$01$'$08.813$''$ in J2000) .  The size of the cross is 4$"$ corresponding to about 70 pc at the adopted distance of 3.8 Mpc to Centaurus A.  All maps except $^{12}$CO(2-1) are at an angular  resolution of 2.25$"$ $\times$ 2.25$"$ while $^{12}$CO(2-1) is at a resolution of 3$"$ $\times$ 3$"$.  The contours of the $^{12}$CO(2-1) map are determined as $5*2^{n/2}$ K km s$^{-1}$ (chosen this way to show the high signal to noise of the map).  The white contour on this map is at roughly 113 K km s$^{-1}$.  The  rare  CO isotopologues all share the common spacing between contours of 0.80 K km s$^{-1}$ corresponding to about 2$\sigma$ in $^{13}$CO(1-0) and 4$\sigma$ in $^{13}$CO(2-1) and C$^{18}$O(2-1).  The first white contours in the $^{13}$CO(2-1), $^{13}$CO(1-0), and C$^{18}$O(2-1) maps are 16.8, 7.2, and 2.4 K km s$^{-1}$ respectively.  The HCO$^{+}$(1-0) integrated intensity map has a contour spacing of 2.1 K km s$^{-1}$ and the first white contour is at a value of 23.1 K km s$^{-1}$.   }
\label{Fig:mom0}
\end{figure*}

\begin{figure*}
\centering
\includegraphics[width=\textwidth]{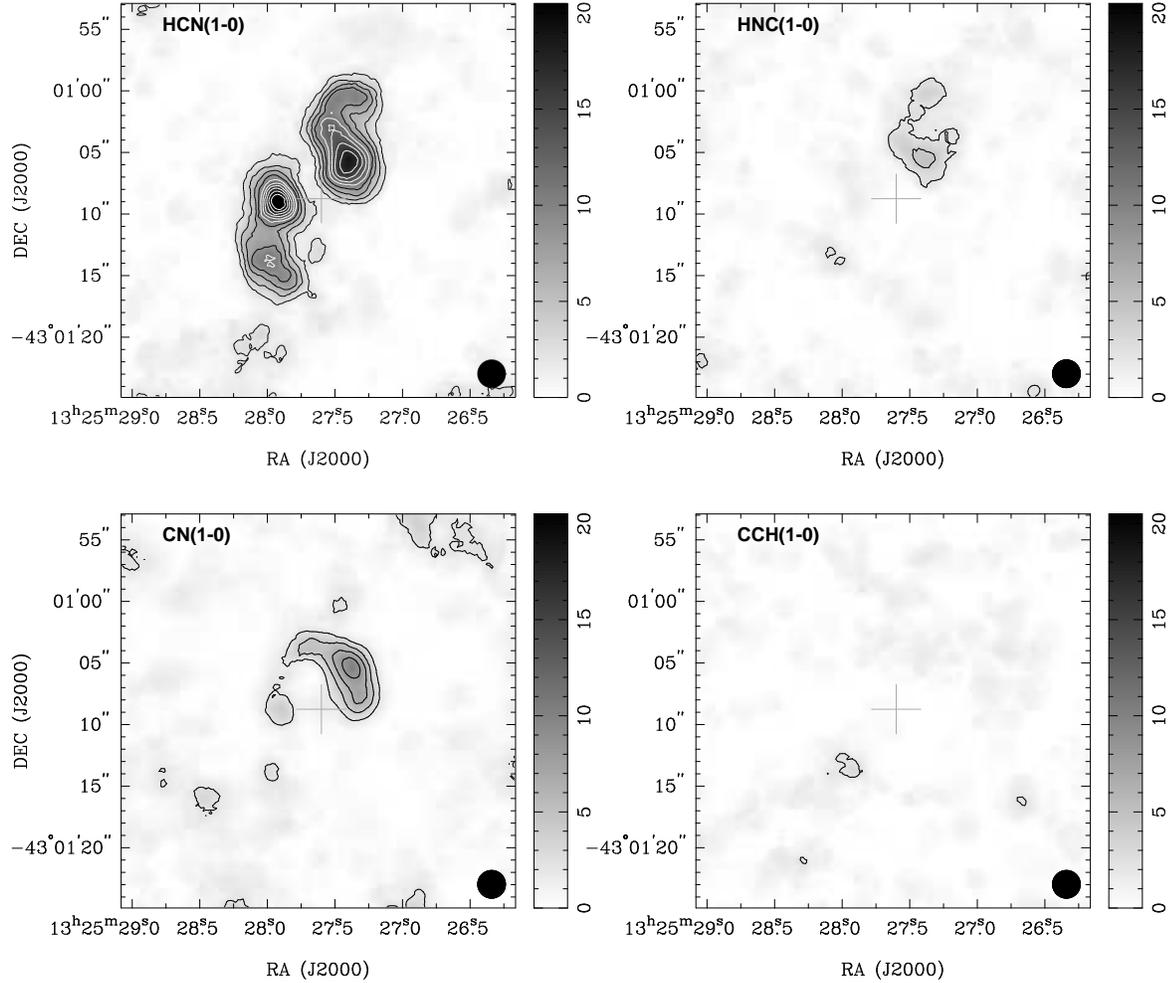}
\caption{ Complementing Figure \ref{Fig:mom0},  integrated intensity maps of the rest of the dense gas tracers are shown here.  The top left panel   displays  HCN($1-0$), top right  displays  HNC($1-0$), bottom left  displays  CN($1-0$) and bottom right  displays  CCH($1-0$).  All of the maps are at a spatial resolution of 2.25$"$ $\times$ 2.25$"$.  The central cross (as in Figure \ref{Fig:mom0}) locates the central black hole and is scaled to 4$"$ or 70 pc at Cen A  (located at RA: 13$^{h}$25$^{m}$27.616$^{s}$	DEC: -43$^{\circ}$01$'$08.813$''$ in J2000) .  All of these images share a contour spacing of 2.1 K km s$^{-1}$ and are displayed on the same colorscale.  This value is approximately 3 $\sigma$ for each of the dense gas tracers. }
\label{Fig:dense}
\end{figure*}

Figure \ref{Fig:mom0} shows the primary beam corrected integrated intensity (moment 0) maps of the detected CO  isotopologues  and HCO$^{+}$ transitions toward the nuclear region of Cen A as well as a comparable $^{12}$CO($2-1$) moment 0 map from ALMA science verification data. Figure \ref{Fig:dense} shows the  other  detected dense gas tracers, HCN($1-0$), CN($1-0$), HNC($1-0$), and CCH($1-0$) toward the same region of Cen A as in Figure \ref{Fig:mom0}.   There are two main structures in the maps visible on our observed scales: the two linear molecular arms, and the dense circumnuclear disk (CND).

In each of the detected CO isotopologue  maps, there are two clumpy linear features, referred to as molecular arms, extending roughly east-west, that coincide with the  parallelogram-shaped  dust lanes.  The two linear features are separated by about 15$"$ (275 pc), each at a projected offset from the SMBH of about 7.5$"$.  Both molecular arms are roughly parallel and at a position angle of  $\sim$ 113$^{\circ}$.    The southern arm is brighter than the northern arm in all detected  molecular lines, with the brightest clumps of the southern arm $\sim 50\%$ brighter than the brightest clumps of the northern arm.  The molecular arms extend beyond the primary beams of both the 3 mm band and 1 mm band observations while maintaining the 15$"$ separation which is consistent with  observations by  \citet{Esp2009}.  

\begin{figure*}
\centering
\includegraphics[width=0.8\textwidth, angle=-90]{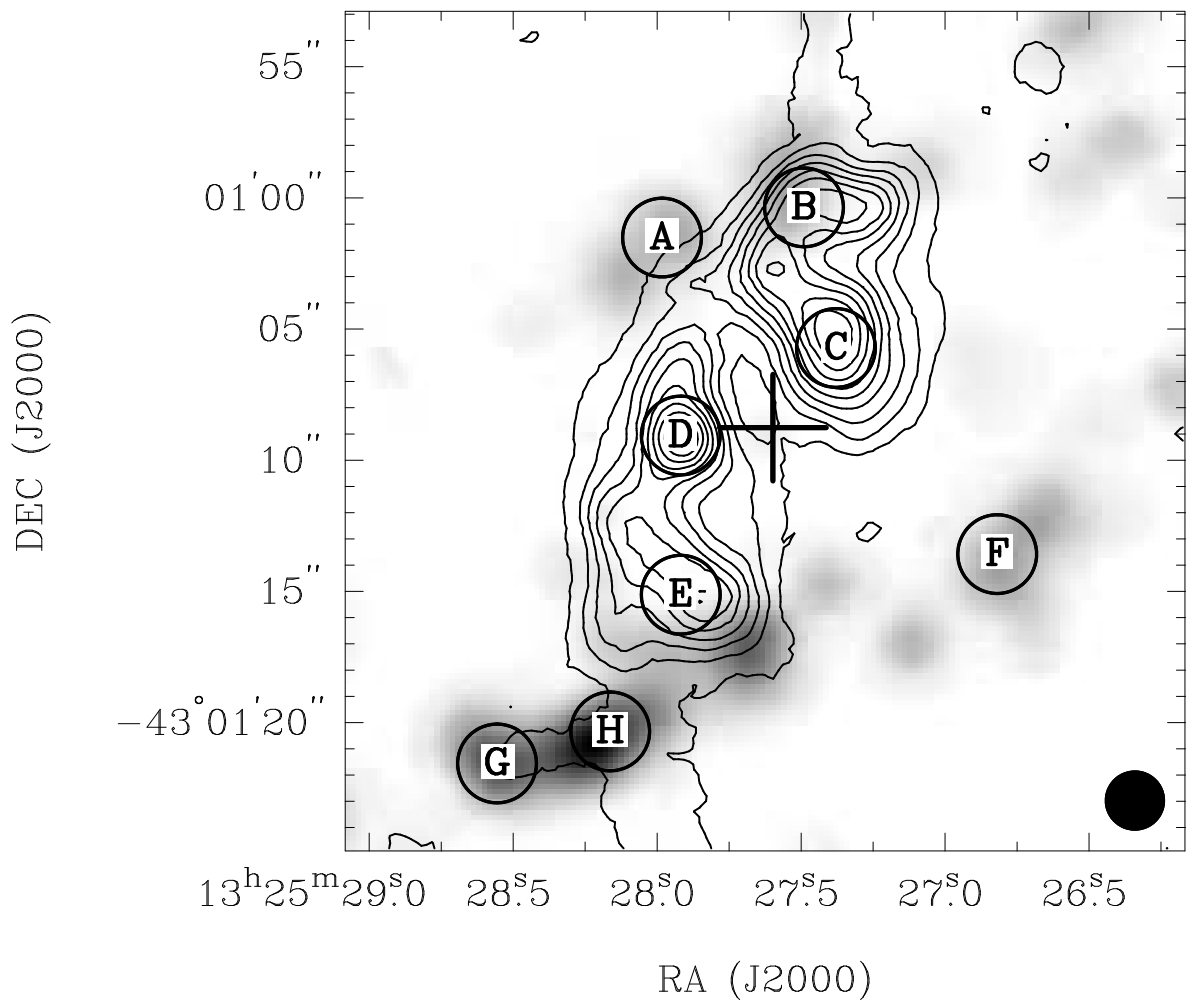}
\caption{A schematic image of the nuclear region of Cen A.  The greyscale in this figure is the integrated intensity map of $^{13}$CO($2-1$) and the contours correspond to the integrated intensity of HCO$^{+}$($1-0$).  The cross locates the central black hole and the lengths of the cross correspond to 4$"$ or 70 pc  (located at RA: 13$^{h}$25$^{m}$27.616$^{s}$	DEC: -43$^{\circ}$01$'$08.813$''$ in J2000) .  The filled circle in the bottom right corner of the figure corresponds to the synthesized beam of 2.25$"$ $\times$ 2.25$"$ common to $^{13}$CO($2-1$) and HCO$^{+}$($1-0$).  The circles labeled {\bf A} through {\bf H} are regions of further study.  Regions {\bf A}, {\bf F}, {\bf G}, and {\bf H} are primarily probing the molecular arms, while regions {\bf C}, {\bf D}, and {\bf E} probe the CND.  Region {\bf B} probes an overlapping region of the molecular arms and the CND.  }
\label{Fig:Schematic}
\end{figure*}

Figure \ref{Fig:Schematic} shows the $^{13}$CO($2-1$) integrated intensity map in greyscale with HCO$^{+}$($1-0$) integrated intensity contours overlaid.  Eight regions are noted in the figure and are used for analysis of the molecular arms and CND.  Region {\bf A}  is  part of  the northern arm, while regions {\bf F}, {\bf G}, and {\bf H} are  located  along the southern molecular arm.  Regions {\bf C}, {\bf D}, and {\bf E} are in the CND.  Region {\bf B} represents a location of overlap between the molecular arms and the CND. The CND is  measured to be $\sim$ 410  pc (22$"$)  $\times$ 215 pc (10.7$"$)  and  the centroid of the CND appears  slighty off-center relative to the AGN ($\sim$ 50 pc). \citet{Gar2014} have observed the molecular CND in  another nearby AGN,  NGC 1068,  to be off-centered from its AGN as well.   

HCN($1-0$), HCO$^{+}$($1-0$), HNC($1-0$), CN($1-0$), and CCH($1-0$) emission are all weak or absent in the molecular arms, but present in the CND.  HCO$^{+}$ traces the majority of the CND, while the others are only detected in specific regions of the CND.  If considered as a thin, flat disk, its axial ratio  of $\sim$2  corresponds to  an inclination angle of $\sim$ 60 degrees and  a  position angle of about 150$^{\circ}$, nearly perpendicular to the radio jet axis.  Thus the minor axis of the CND is at a position angle of 60$^{\circ}$.  The FWHM  (full width at half maximum)  linewidth of the CND is around 40 km s$^{-1}$ for an inner diameter of 140 pc corresponding to {\bf C} and {\bf D} (Figure \ref{Fig:Schematic}) and 70 km s$^{-1}$ between diameters of 275 and 325 pc corresponding to {\bf B} and {\bf E} (Figure \ref{Fig:Schematic}) measured by HCO$^{+}$($1-0$).  The outer portions of  the  CND for these transitions are coincident with the edges of the CND that are seen in $^{13}$CO($2-1$).   Regions {\bf C} and {\bf D} are most likely composed of a combination of nuclear filaments and the nuclear ring structure \citep{Esp2017}. 

In the $^{12}$CO($2-1$) and $^{13}$CO($2-1$) images, the outer edge of the CND is also detected, however no part of the CND is detected in either $^{13}$CO($1-0$) or C$^{18}$O($2-1$).   The clumps  in  the molecular arms have  spectral line widths (FWHM)  of $\sim 15$ km s$^{-1}$ while the detected portions of the CND have FWHM of $\sim 40$ km s$^{-1}$  as measured by $^{13}$CO($2-1$) (determined in Section \ref{spectra}), which are consistent with the measured line widths from the dense gas tracers. 

The CND is considerably brighter along its major axis than its minor axis.  The minor axis at a position angle of approximately 60$^{\circ}$ separates regions {\bf B} and {\bf C} on the northern side from {\bf D} and {\bf E} on the southern side.  From the HCO$^{+}$($1-0$) emission, the southern side of the minor axis is almost not present, and the northern side of the minor axis is weaker than any of the emission along the CND's major axis.  The underluminous minor axis may be due to deviations from a uniform disk and the underluminous minor axis was also observed  in $^{12}$CO($2-1$)  with the SMA as reported  by  \citet{Esp2009}.  

\subsection{Molecular Gas  Kinematics} \label{Sec:Kinem}

\begin{figure*}
\centering
\includegraphics[width=0.7\textwidth]{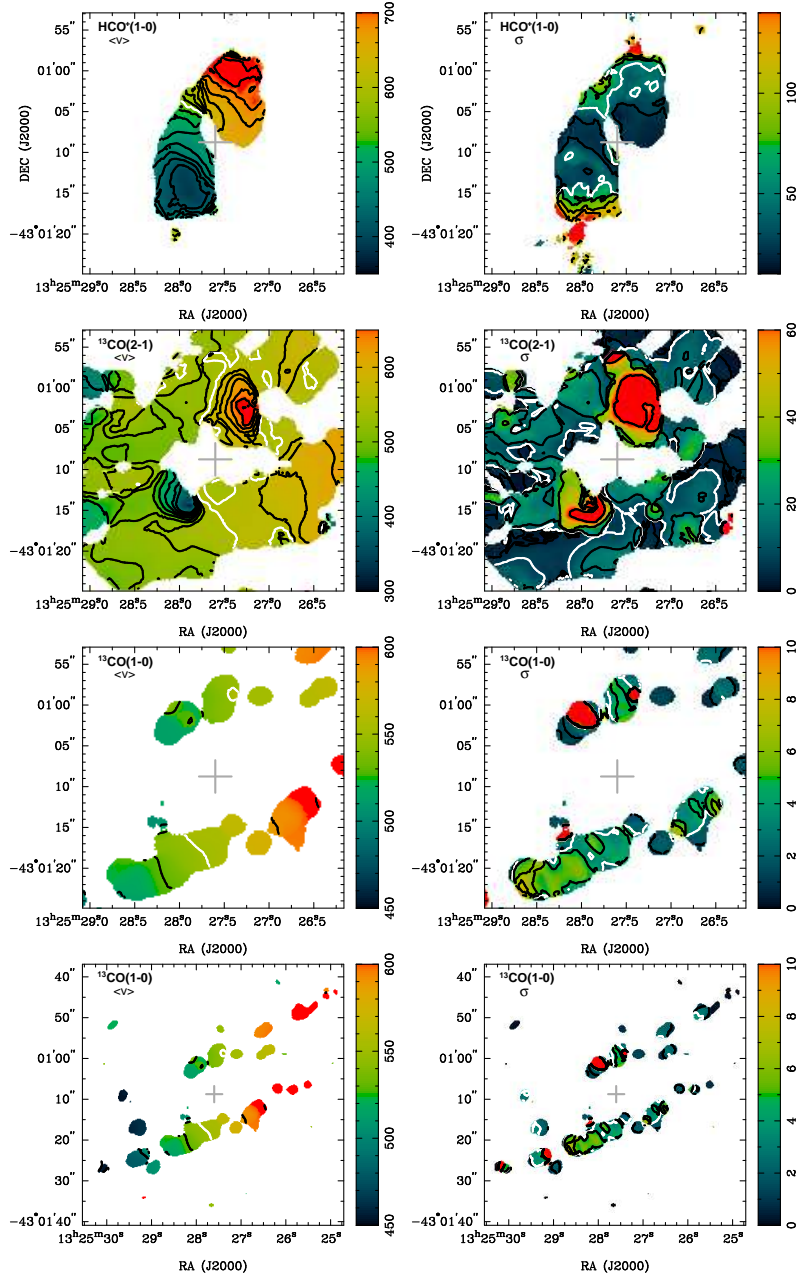}
\caption{The intensity weighted velocity field (moment 1) [Left] and the intensity weighted velocity dispersion (moment 2) [Right] are shown here  for HCO$^{+}$($1-0$), $^{13}$CO($2-1$), and $^{13}$CO($2-1$).  The last image in each  column  is a full field view of the $^{13}$CO($1-0$) moment 1 map showing the entire  3 mm  primary beam.  Each of the moment 1 maps  has  a contour spacing of 20 km s$^{-1}$.  The white contour in each of these images corresponds to a velocity of 550 km s$^{-1}$.  The HCO$^{+}$(1-0) moment 2 map has a contour spacing of 25 km s$^{-1}$ and the white contour marks a velocity dispersion of 50 km s$^{-1}$.  The moment 2 map of $^{13}$CO(2-1) has contour spacing of $2^{n}$ km s$^{-1}$ with n=1,2,3,4,5,6 (chosen to show the wide range of velocity dispersion).  The white contour in this image is at 8 km s$^{-1}$.  The $^{13}$CO($1-0$) moment 2 map has a contour spacing of 1.5 km s$^{-1}$ with the white contour marking 3 km s$^{-1}$.  }
\label{Fig:VelDisp}
\end{figure*}

The intensity-weighted mean velocity (moment 1) and intensity-weighted velocity dispersion (moment 2) maps are shown in Figure \ref{Fig:VelDisp}, for the HCO$^{+}$($1-0$), $^{13}$CO($1-0$), and $^{13}$CO($2-1$) transitions.  The HCO$^{+}$($1-0$)  emission  traces  the CND very well in both velocity and intensity but  there is no emission from  the arms. The $^{13}$CO($1-0$)  emission  on the other hand,  traces  the arms  but  not   the CND.  $^{13}$CO($2-1$) exhibits both the molecular arms and the CND components and resembles a combination of the HCO$^{+}$($1-0$) and $^{13}$CO($1-0$) velocity maps.  The $^{13}$CO($1-0$) transition covers velocities of $\sim$ 500 -- 600 km s$^{-1}$,  and shows  a velocity gradient of $\sim$ 0.2 km s$^{-1}$ pc$^{-1}$  increasing from east to west,  with a velocity dispersion of $<$ 10 km s$^{-1}$.  HCO$^{+}$($1-0$) covers the velocity range from about 330 to 770 km s$^{-1}$,  reveals a  much steeper velocity gradient along the same axis  of 1.2 km s$^{-1}$ pc$^{-1}$, with a dispersion of $\sim$ 50 km s$^{-1}$.  Using the central velocity of the CND as  observed  by HCO$^{+}$($1-0$), we calculate that the systemic velocity  in the LSRK (Local Standard of Rest, kinematic definition) velocity frame   for Cen A is approximately 550 km s$^{-1}$, which is consistent with the values in the literature which range from about 530 -- 560 km s$^{-1}$ \citep{Qui1992,Mar2001,Hae2006}.

Figure \ref{Fig:VelDisp}  also shows  the full field $^{13}$CO($1-0$) intensity weighted mean  velocity and velocity dispersion maps.  These maps show that outside the field limited by the $^{13}$CO($2-1$) primary  beam, the $^{13}$CO($1-0$) continues to follow a similar trend of lower  dispersion and a smaller  velocity gradient than HCO$^{+}$($1-0$).  

The molecular arm component in the $^{13}$CO($2-1$) moment 1 map ranges  from 450 km s$^{-1}$ up to 600 km s$^{-1}$ [similar to the range of $^{13}$CO($1-0$)] while the circumnuclear component of the $^{13}$CO($2-1$) map covers 375 km s$^{-1}$ to 675 km s$^{-1}$ [similar to the range of HCO$^{+}$($1-0$)].  There are two locations in  the map, near {\bf B} and {\bf E},  where the dispersion jumps to $>$ 60 km s$^{-1}$. , but the dispersion remains  around $\sim 10$ km s$^{-1}$  across the remaining map.   This jump in the dispersion is not due to a single component with a dispersion of 60 km s$^{-1}$, but the positional overlap of the narrow dust lane feature with the broad circumnuclear feature.  It is worth noting that the velocity field of the inner region of the CND seems to twist in the  HCO$^{+}$($1-0$)  moment 1 map (Figure \ref{Fig:VelDisp}).  This warping appears to decrease the position angle of the inner CND, consistent  with the H$_{2}$ velocity map of \citet{Neu2007}  and CO maps of \citet{Esp2017}.   However, in the H$_{2}$ velocity map, there appears to be another warp further in, not sampled by our data  (due to the absorption).   

\begin{figure*}
\centering
\includegraphics[width=\textwidth]{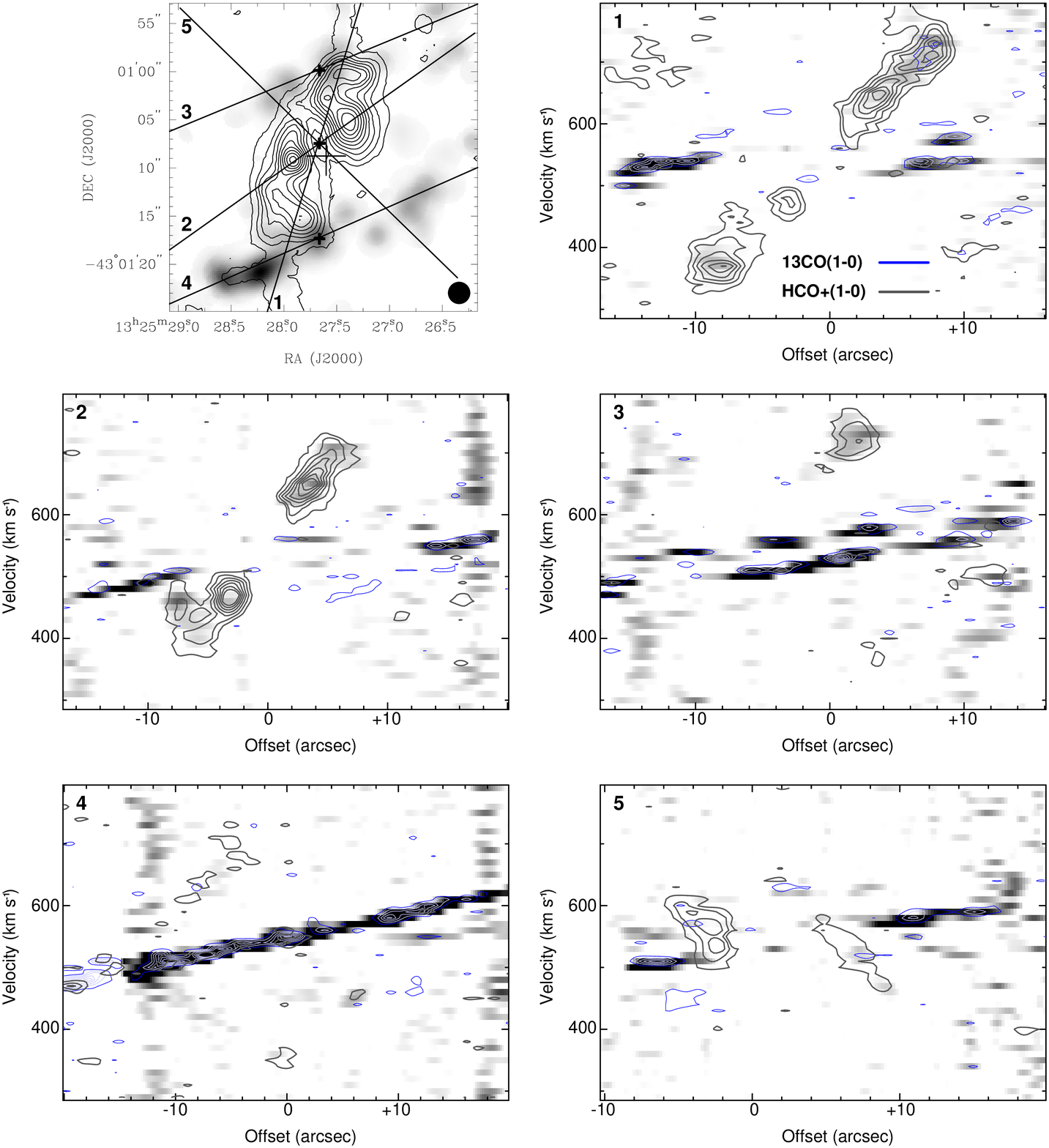}
\caption{This figure shows position-velocity (PV) diagrams for the inner 500 pc  of Cen A.  The top-left panel shows the locations of the slices taken to make the PV diagrams.  Each cut is taken from east to west.  The greyscale in all of the images corresponds to $^{13}$CO($2-1$).  The black/grey contours in each of the images correspond to HCO$^{+}$($1-0$) while the blue contours (absent in the top-left panel) show $^{13}$CO($1-0$).  The large cross in the top-left panel marks the location of the central black hole, while the smaller  black  crosses mark the  zero offset  of each of the slices ($1'' \simeq 18$ pc offset). Slices 1, 2, and 5 all share the same central point.   }
\label{Fig:PVdiag}
\end{figure*}

Position-Velocity (PV) diagrams for $^{13}$CO($2-1$), $^{13}$CO($1-0$), and HCO$^{+}$($1-0$) are shown in Figure \ref{Fig:PVdiag}.  The top left  panel  of the figure shows  the locations of the slices through the cube.  Five slices were taken; the first through the major axis of the CND, the second through the inner peaks of the CND, the third and fourth are along the northern and southern molecular arms, and the last is along the axis of the radio jet.  The slices were all averaged over a width of 11 pixels, approximately 40 pc.  These pixels are not independent and were chosen such that the width of the slice was approximately the same width as the beam.   The broad components in all of the PV diagrams come from the CND, while the narrow components come from the molecular arms.  The center of the slice (the 0 offset location) is noted by a small cross on the slice line.  Slices 1, 2, and 5 all share the same center.

Slice 1 extends along the elongated axis of the CND while also crossing the molecular arms.  Slice 1 shows the lack of $^{13}$CO($1-0$) along with the prominence of HCO$^{+}$ in the CND.  $^{13}$CO($2-1$) is present in both the molecular arm components as well as the CND components, although not as prominently as HCO$^{+}$.  The CND has components spanning a velocity range from 350 -- 800 km s$^{-1}$.  The molecular arms span a much narrower velocity range, about 520 -- 580 km s$^{-1}$ within the band 6 primary beam in Slice 1.  

Slice 2 intersects the inner portion of the nuclear disk and avoids the molecular arms until large offsets.  Again, there is a lack of $^{13}$CO($1-0$) in the broad (CND) components.  $^{13}$CO($2-1$) is present in both the molecular arms as well as the CND although more prominent in the molecular arms.  This slice shows that the inner portion of the CND has a velocity range of 375 -- 700 km s$^{-1}$ while  both  molecular arms span velocities  $\sim$ 475 -- 575 km s$^{-1}$.  

Slices 3 and 4 primarily cover the northern and southern  molecular arms, crossing only the very edges of the CND.  Both slices show the narrow velocity range of the molecular arms, but slice 4  (southern arm)  is simply a linear feature in the PV diagram.  Slice 3 has a more complicated structure.  Slice 3  (northern arm) contains a linear feature similar to slice 4 with the addition of $^{13}$CO($1-0$) and $^{13}$CO($2-1$) detected in the same position as the linear feature, but offset by $\sim$ 30 km s$^{-1}$  showing a double structure. This double structure might be a result of overlapping tilted ring along the line of sight, but its origin is not entirely evident.    The absence of emission from $^{13}$CO(2-1) at an offset of $-$14$"$ to $-$20$"$ results from being outside the 20\% level of the primary beam.

Slice 5 traces approximately the AGN jet axis.  However, there is no indication of significant structure that would come from an interaction of the molecular gas with the jet.  
  
\subsection{Spectra}

\subsubsection{Emission Spectra} \label{spectra}

\begin{figure*}
\centering
\includegraphics[width=\textwidth]{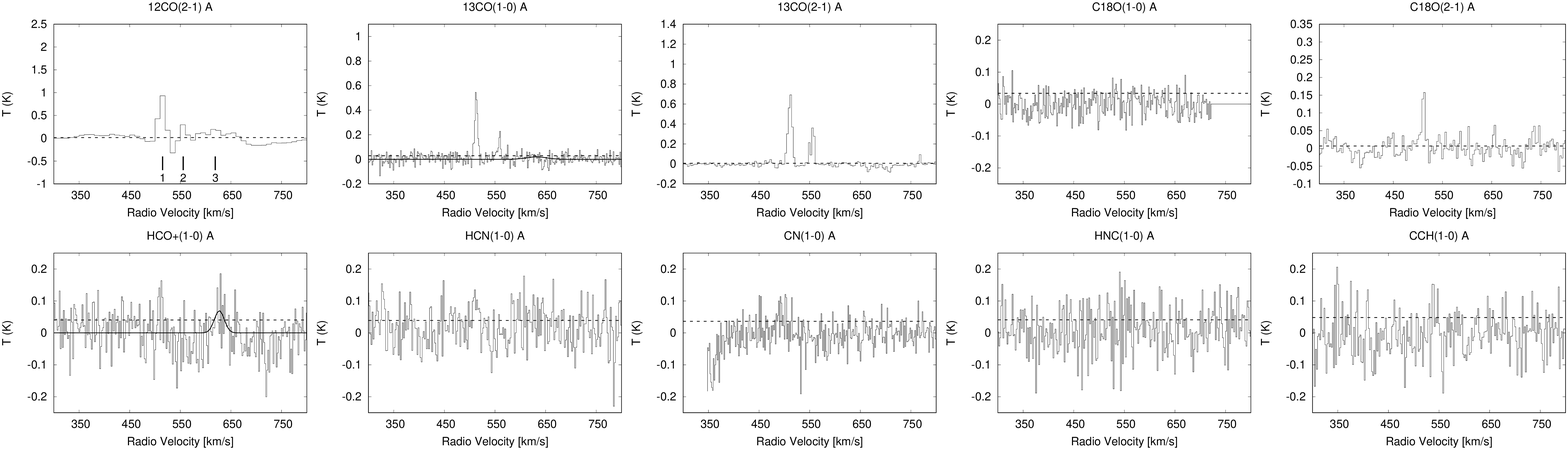}
\caption{Emission spectra taken  toward  region A  (See Figure \ref{Fig:Schematic};  RA: 13$^{h}$25$^{m}$27.983$^{s}$ DEC: -43$^{\circ}$01$'$01.519$''$ in J2000) .  In this and figure \ref{Spec:B} through \ref{Spec:H}: the top row shows (from  left to right ) $^{12}$CO(2-1), $^{13}$CO(1-0), $^{13}$CO(2-1), C$^{18}$O(1-0), and C$^{18}$O(2-1).  The bottom row shows (from  left to right ) HCO$^{+}$(1-0), HCN(1-0), CN(1-0), HNC(1-0), and CCH(1-0). The velocity  axes range from 300 to 800 km s$^{-1}$ and the  brightness  temperature  axes are  adjusted for each plot to highlight  the emission.  The dashed line on each panel of this figure shows the 1 $\sigma$ noise level for that transition.  For Figures \ref{Spec:B} -- \ref{Spec:H} the corresponding transition will have the same 1 $\sigma$ limit. For this figure as well as Figures \ref{Spec:B} -- \ref{Spec:H}, each detected feature that is $\leq 3 \sigma$ the Gaussian fit is shown as a solid line to guide the eye.    }
\label{Spec:A}
\end{figure*}

\begin{figure*}
\centering
\includegraphics[width=\textwidth]{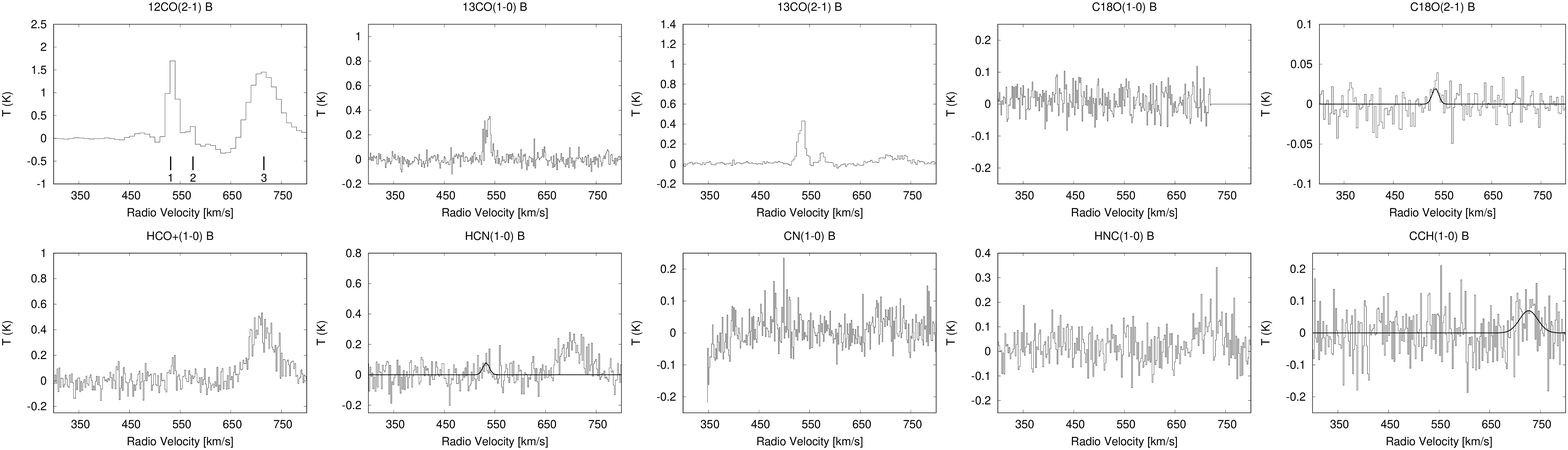}
\caption{Emission spectra taken  toward  region B  (See Figure \ref{Fig:Schematic};  RA: 13$^{h}$25$^{m}$27.490$^{s}$	DEC: -43$^{\circ}$01$'$00.370$''$ in J2000).  See Figure \ref{Spec:A} for details.} 
\label{Spec:B}
\end{figure*}

\begin{figure*}
\centering
\includegraphics[width=\textwidth]{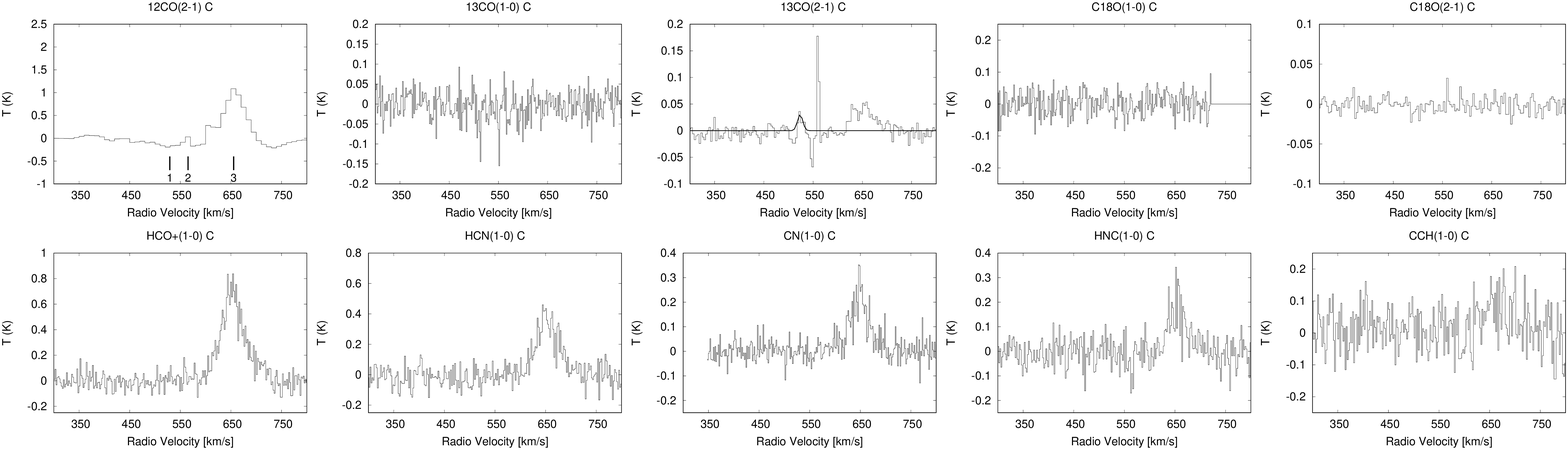}
\caption{Emission spectra taken  toward  region C  (See Figure \ref{Fig:Schematic};  RA: 13$^{h}$25$^{m}$27.379$^{s}$	DEC: -43$^{\circ}$01$'$05.728$''$ in J2000).  See Figure \ref{Spec:A} for details.} 
\label{Spec:C}
\end{figure*}

\begin{figure*}
\centering
\includegraphics[width=\textwidth]{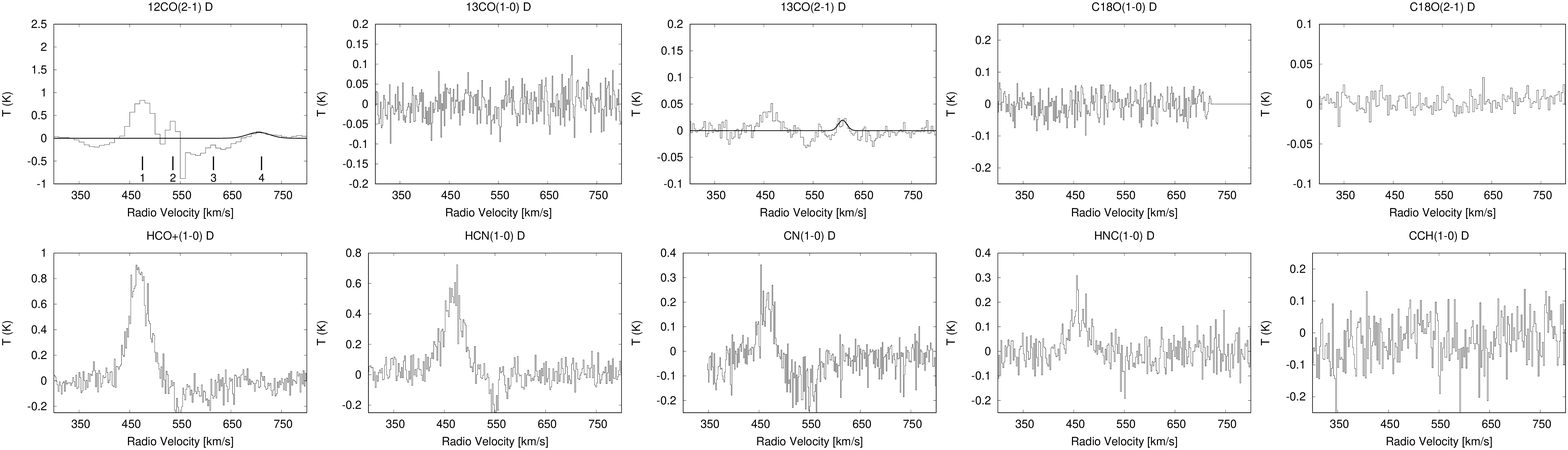}
\caption{Emission spectra taken  toward  region D  (See Figure \ref{Fig:Schematic};  RA: 13$^{h}$25$^{m}$27.918$^{s}$	DEC: -43$^{\circ}$01$'$09.025$''$ in J2000).  Being near the AGN, there are absorption artifacts introduced during the continuum subtraction. See Figure \ref{Spec:A} for details.} 
\label{Spec:D}
\end{figure*}

\begin{figure*}
\centering
\includegraphics[width=\textwidth]{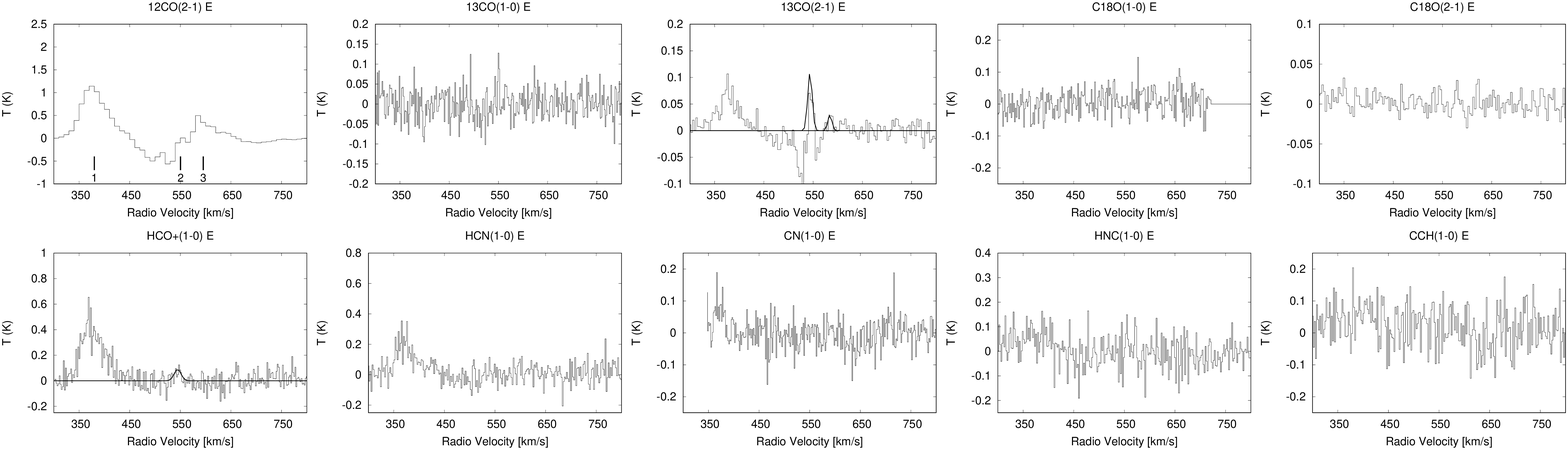}
\caption{Emission spectra taken  toward  region E  (See Figure \ref{Fig:Schematic};  RA: 13$^{h}$25$^{m}$27.918$^{s}$	DEC: -43$^{\circ}$01$'$15.132$''$ in J2000).  Being near the AGN, there are absorption artifacts introduced during the continuum subtraction. See Figure \ref{Spec:A} for details.} 
\label{Spec:E}
\end{figure*}

\begin{figure*}
\centering
\includegraphics[width=\textwidth]{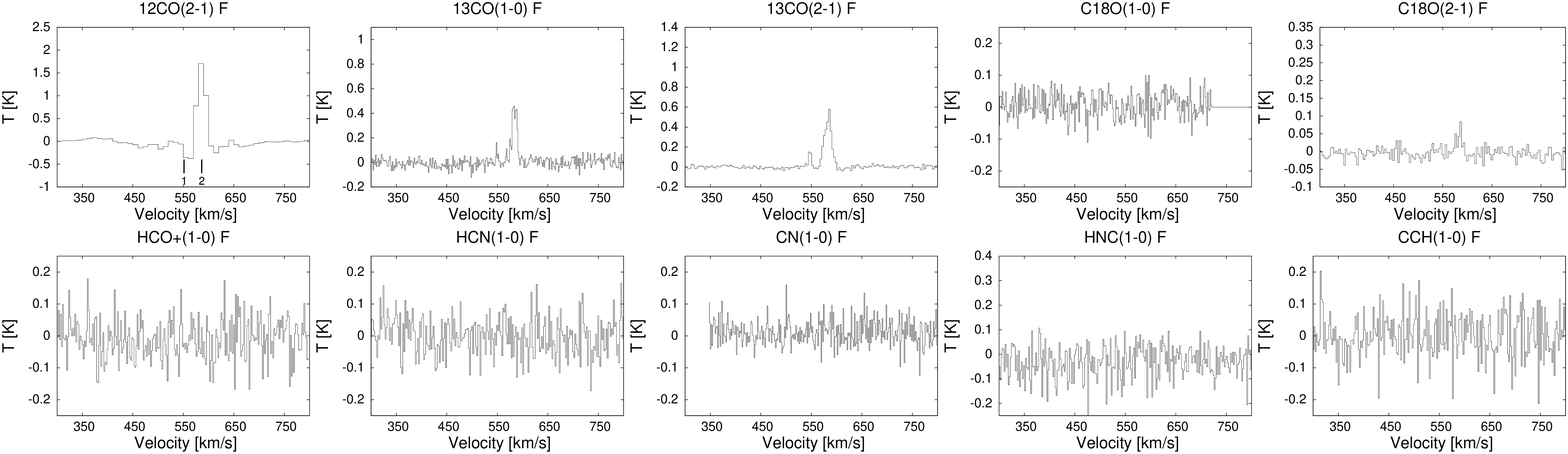}
\caption{Emission spectra taken  toward  region F  (See Figure \ref{Fig:Schematic};  RA: 13$^{h}$25$^{m}$26.819$^{s}$	DEC: -43$^{\circ}$01$'$13.585$''$ in J2000).  See Figure \ref{Spec:A} for details.} 
\label{Spec:F} 
\end{figure*}

\begin{figure*}
\centering
\includegraphics[width=\textwidth]{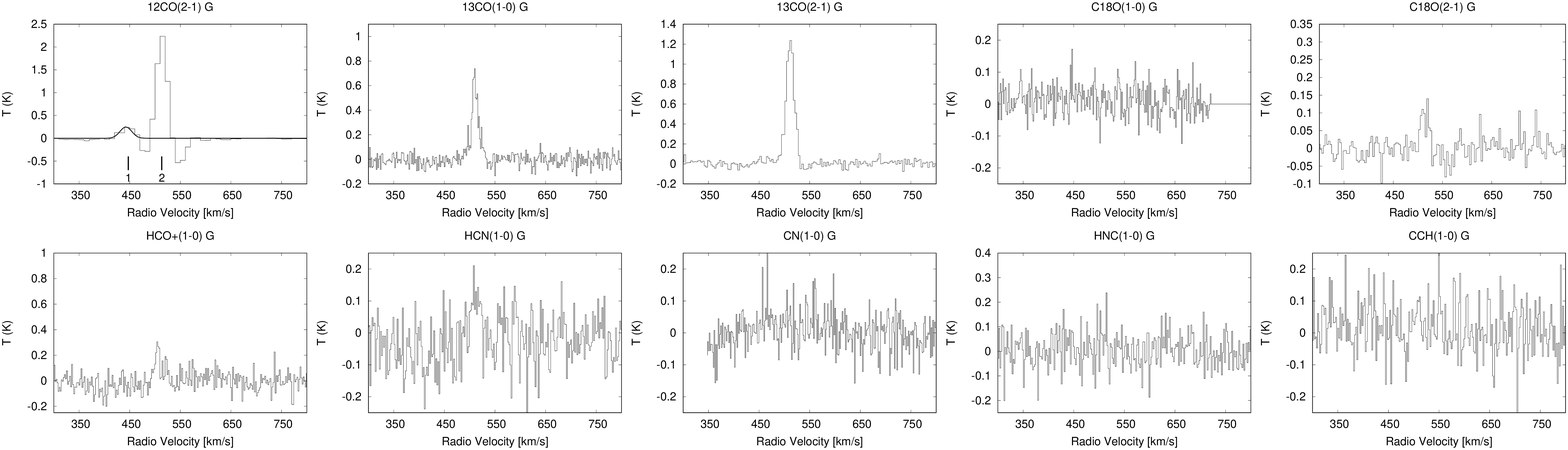}
\caption{Emission spectra taken  toward  region G  (See Figure \ref{Fig:Schematic};  RA: 13$^{h}$25$^{m}$28.556$^{s}$	DEC: -43$^{\circ}$01$'$21.565$''$ in J2000).  See Figure \ref{Spec:A} for details.} 
\label{Spec:G}
\end{figure*}

\begin{figure*}
\centering
\includegraphics[width=\textwidth]{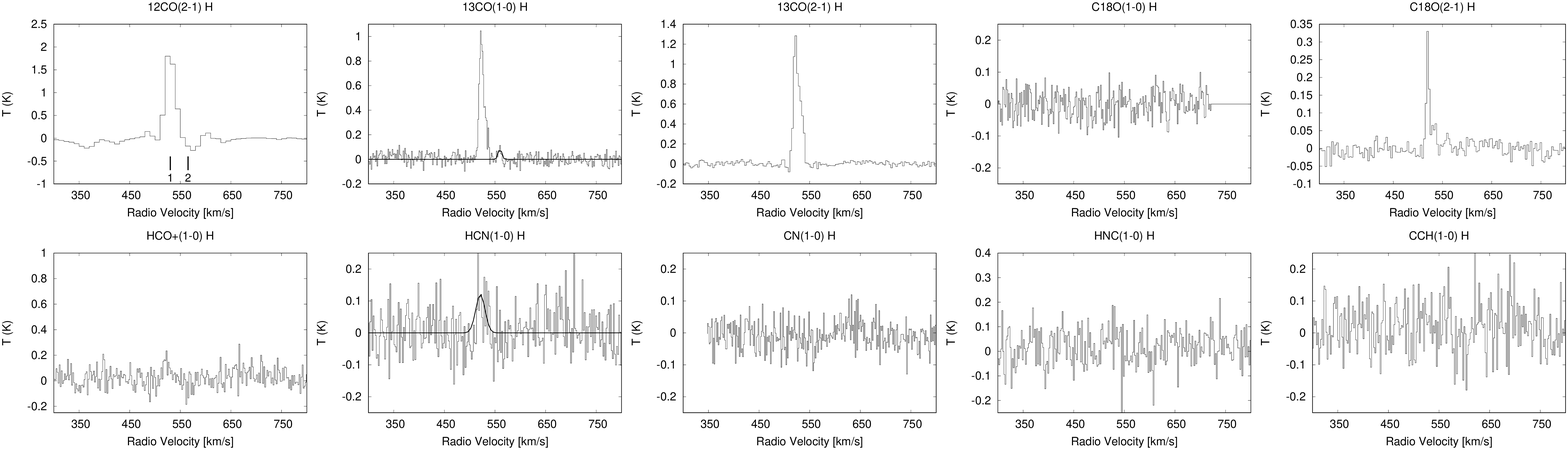}
\caption{Emission spectra taken  toward  region H  (See Figure \ref{Fig:Schematic};  RA: 13$^{h}$25$^{m}$28.556$^{s}$	DEC: -43$^{\circ}$01$'$20.343$''$ in J2000).  See Figure \ref{Spec:A} for details.} 
\label{Spec:H}
\end{figure*}

 Figures \ref{Spec:A} to \ref{Spec:H} show the spectra from each of the detected transitions as well as $^{12}$CO($2-1$) from the science verification data and the non-detection of C$^{18}$O($1-0$) toward the eight marked locations in Fig. \ref{Fig:Schematic}. Spectra were taken using a 2.25'' apertures corresponding to the circles in Figure \ref{Fig:Schematic}.    We identify 22 velocity components throughout all of the observed molecular transitions and spatial locations. Each of these components are labeled by  its  spatial location and a numeric identifier that increases with velocity, irrespective of the transition. These components are identified in the top left panel in Figures \ref{Spec:A} through \ref{Spec:H} (the $^{12}$CO($2-1$) spectra from each of the regions).  Each figure shows the spectra from each of the detected transitions as well as $^{12}$CO($2-1$) from the science verification data and the non-detection of C$^{18}$O($1-0$). The properties of these spectral features, i.e. for each velocity component (e.g. {\bf A-1} , {\bf A-2}, {\bf A-3}),  are given in Table \ref{tab:lw} and corresponding line ratios are given in Table \ref{tab:ratio}.  For undetected features, a one sigma  upper  limit for the peak temperature is provided.  Comparisons between detections and non-detections in Table \ref{tab:ratio} are presented as upper or lower limits of that line ratio.  Comparisons between  double non-detections are not presented in   Table \ref{tab:ratio}.

Regions {\bf A}, {\bf F}, {\bf G}, and {\bf H} represent the molecular arm components.  These regions typically contain a couple of features that have narrow linewidths on the order of 10 -- 15 km s$^{-1}$.  These features are prominent in CO and its isotopologues, but nearly non-existent in the dense gas tracers.  HCO$^{+}$($1-0$) and HCN($1-0$) show faint  narrow  features in regions {\bf B}, {\bf G}, and {\bf H}. In regions {\bf B} and {\bf G}, HCO$^{+}$($1-0$) and HCN($1-0$) are  at  about 1/2 and 1/4 of the corresponding $^{13}$CO($1-0$) peak temperature  values, and in region {\bf H} they are about 1/6 and 1/8 the $^{13}$CO($1-0$) value.  In region {\bf A} there is a feature near 615 km s$^{-1}$ present in  three  of the transitions that have broad linewidths, comparable to features in the CND.  This feature corresponds to emission from the edge of the CND contributing to  region {\bf A}.  

Regions {\bf C}, {\bf D}, and {\bf E} show the emission  primarily  from the CND.  These regions contain spectral features that are prominent in the dense gas tracers, and are nearly absent in CO.  $^{12}$CO($2-1$) and $^{13}$CO($2-1$) both contain these features from the CND, but the features are not seen in $^{13}$CO($1-0$) or either of the C$^{18}$O transitions.  These features all have much broader linewidths than the features on the molecular arms, with widths on the order of 50 km s$^{-1}$.  

Region {\bf B} contains spectral features attributed to the northern  molecular arm as well as the CND.   This is probably best seen in the $^{12}$CO($2-1$) and $^{13}$CO($2-1$) spectra.  Components {\bf B-1} and {\bf B-2} have narrow linewidths and closely resemble the features of regions {\bf A}, {\bf F}, {\bf G}, and {\bf H}, which correspond to the molecular arms.  Component {\bf B-3} has a much broader linewidth more similar to the features in regions {\bf C}, {\bf D}, and {\bf E}.  Region {\bf B}, therefore, shows a spatial region where the molecular arm component and the CND overlap.  

\subsubsection{Absorption Spectra} \label{abssec}

\begin{figure*}
\centering
\includegraphics[width=\textwidth]{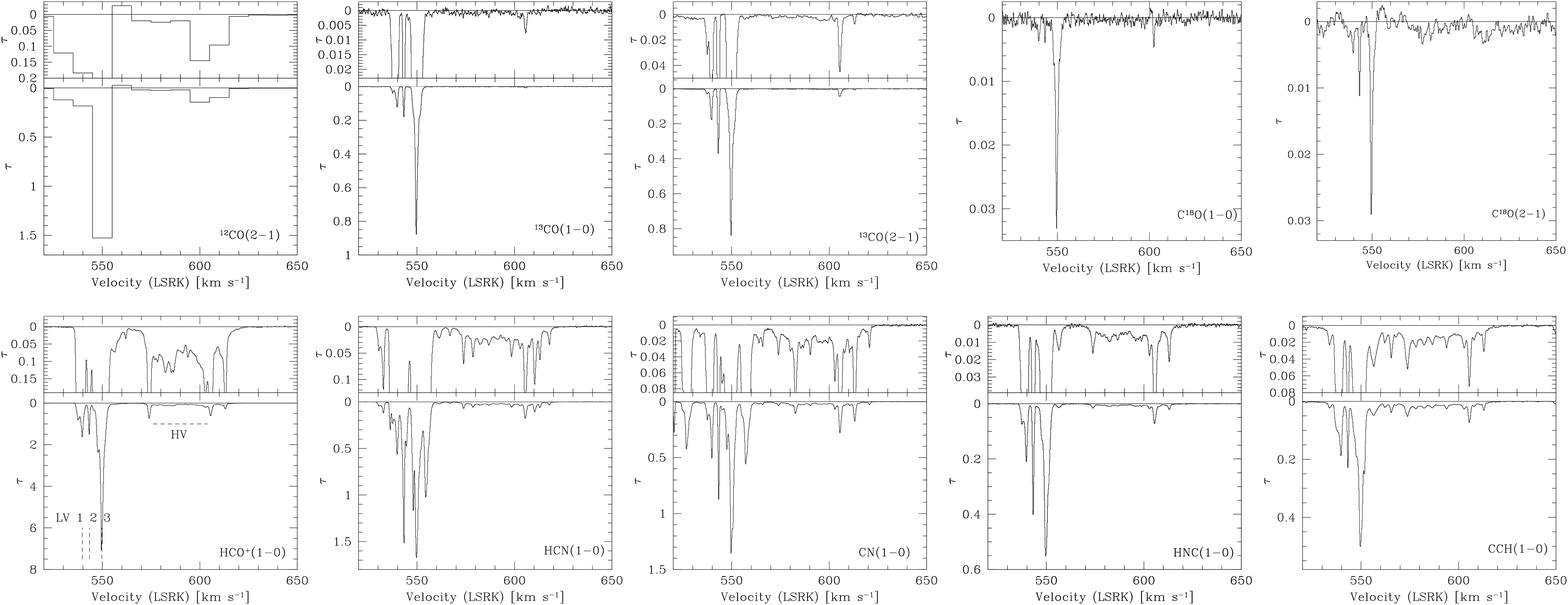}
\caption{The absorption spectra taken toward the core of Cen A. The ordering of the top and bottom rows are the same as in Figures \ref{Spec:A} through \ref{Spec:H}.  For all transitions except the two C$^{18}$O  lines , a second panel above the full spectrum focuses on the lower level high velocity absorption component and the  spectral  noise. HCN($1-0$), CN($1-0$) and CCH($1-0$) all have hyperfine structure, but we only consider the peak of the main hyperfine component.  }
\label{Spec:abs}
\end{figure*}

The absorption spectra (Figure \ref{Spec:abs}) were made by taking a spectrum through the central point of an image cube for each transition.    Each spectrum was then converted to units of optical depth ($\tau$) based on the continuum flux from Table \ref{tab:obs} assuming a continuum source covering factor of unity  (see \citealt{Mul2009}, their Section 4.1 for the size scale of the continuum)  and the equation $\tau = - \ln[1-(\frac{-\mathrm{F_{spec}}}{\mathrm{F_{cont}}})]$, where $\mathrm{F_{spec}}$ and $\mathrm{F_{cont}}$ are the continuum subtracted line flux  and the continuum flux, respectively.  There are two groups of absorption lines detected in the various molecular lines.  There is one group near the systemic velocity of Cen A  (550 km s$^{-1}$, Section \ref{Sec:Kinem}) that has three  main components within 10 km s$^{-1}$  and multiple blended or weaker components called the low velocity complex (LV complex).  The other group is redshifted by 20 -- 70 km s$^{-1}$ and contains many blended components called the high velocity complex (HV complex).   Both the low and high velocity complexes were previously detected  in  HCO$^{+}$($1-0$), HCN($1-0$), and HNC($1-0$) at a lower velocity resolution and sensitivity  by  \citet{Wik1997}.  We restrict our analysis of the absorption spectra to three of the narrow components (539.7, 543.3, and 549.7 km s$^{-1}$) in the LV complex and a median value taken from 576 -- 604 km s$^{-1}$ of the HV complex and only include the molecular transitions  C$^{18}$O($1-0$)  (where no emission was detected)  and those  also  detected in emission.  

\section{Discussion}

The flow of gas in the vicinity of an AGN is important in understanding the fueling of the AGN.  The nuclear region of Cen A has multiple components of molecular gas, including two linear features that cross nearly in front of the AGN (See $^{13}$CO($1-0$) panels in Figure \ref{Fig:PVdiag}), a circumnuclear disk approximately 400 pc in diameter, and absorption complexes both near  the  systemic  velocity  (LV)  and redshifted by 20 -- 70 km s$^{-1}$  (HV).  We investigate connections between these components both dynamically and chemically.

\subsection{ Are the Molecular Arms and the CND Connecting? }

We do not detect emission connecting the narrow linewidth emission components of the molecular arms, and the broad linewidth emission from the molecular CND.  Region {\bf B} and partially {\bf E} in Figure \ref{Fig:Schematic} mark the locations where the CND and the molecular arms appear to  spatially  overlap. According to the moment 2 maps of $^{13}$CO($2-1$), regions {\bf B} and {\bf E} correspond to high velocity dispersion regions.  However, looking at the spectra  taken toward these regions  (See Figures \ref{Spec:B} and \ref{Spec:E}), the cause of the high dispersion is multiple separate kinematic components along the lines  of sight.  In region {\bf B} there is a weak broad component and two narrow components that are kinematically distinct.  In region {\bf E} there is a  single narrow component and a kinematically distinct broad component.  In the PV diagrams for slices 1 and 2, (Figure \ref{Fig:PVdiag}), there does not appear to be any connection between the broad, high velocity gradient component, corresponding to the CND, and the narrow, low velocity gradient component, corresponding to the molecular arms. However, due to the missing short spacings of the observations, there may be large scale structure that is resolved out possibly hiding the connection  described  in \citet{Esp2009}. An absence of an observed  connection between the  two main emission components  means the method for transporting gas from the  molecular arms  to the  CND  is still not clear.

\subsection{Gas Chemistry}
 
\subsubsection{CO Isotopic Ratios} \label{Sec:Isotope}

 Intensity ratios between isotopologue transitions with the same J states, I[$^{12}$CO($2-1$)]/I[$^{13}$CO($2-1$)], I[$^{12}$CO($2-1$)]/I[C$^{18}$O($2-1$)], and I[$^{13}$CO($2-1$)]/I[C$^{18}$O($2-1$)] contain information on the opacities and abundance ratios of CO and its isotopologues, assuming each pair of transitions  has  a common excitation temperature  T$_{x}$.  For example, the intensity ratio, I[$^{12}$CO($2-1$)]/I[$^{13}$CO($2-1$)], with the same line width for 2-1 and 1-0 and in the absence of fractionation or selective photodissociation,  gives:
 
\[
 \frac{\rm{^{12}T_{21}}}{\rm{^{13}T_{21}}} \simeq \frac{^{12}f_{a}(^{12}J_{21}(T_{x})- ^{12}J_{21}(T_{bg}))(1-e^{-^{12}\tau})}{^{13}f_{a}(^{13}J_{21}(T_{x})- ^{13}J_{21}(I_{bg}))(1-e^{-A^{12}\tau})},
\]
 
\noindent with $^{13}\tau$ replaced by $A^{12}\tau$, where A is the isotopic abundance ratio [$^{12}$CO/$^{13}$CO] and with:

\[
J_{\nu}(T) = \frac{\frac{\rm{h}\nu}{\rm{k}}}{e^{\frac{\rm{h}\nu}{\rm{k}T_{x}}} - 1}.
\]

\noindent In the high temperature limit, $^{12}J_{ij}(T_{ij})$ simply becomes $T_{ij}$ and assuming similar  area  filling factors, $^{i}f_{a}$, and the same T$_{x}$,  this can be simplified to  $^{12}$I$_{21}$/$^{13}$I$_{21} = (1-e^{-^{12}\tau})/(1-e^{-A^{12}\tau})$ (e.g. \citealt{Aal1995}).
 
 Similarly, taking ratios of the rotational transitions of $^{13}$CO (and C$^{18}$O),  between different J states of the same isotopologue like I[$^{13}$CO($2-1$)]/I[$^{13}$CO($1-0$)], constrain T$_{x}$. Again, assuming high temperature limit and identical line widths and filling factors, the line temperature ratio is approximated by the equation:
 
\[ \frac{\rm{^{13}T^{21}_{mb}}}{\rm{^{13}T^{10}_{mb}}} \simeq \frac{(\rm{T_{21}-T_{21}^{bg}})(1-e^{-^{13}\tau_{21}})}{(\rm{T_{10}-T_{10}^{bg}})(1-e^{-^{13}\tau_{10}})}.
\]
 
 \noindent In the optically thick limit this becomes:
 
\[ \frac{\rm{^{13}T^{21}_{mb}}}{\rm{^{13}T^{10}_{mb}}} \simeq \frac{(\rm{T}_{21}-\rm{T}_{21}^{bg})}{(\rm{T}_{10}-\rm{T}_{10}^{bg})} \simeq 1,
\]
 
\noindent and in the optically thin limit  we obtain : 

\[ \frac{\rm{^{13}T^{21}_{mb}}}{\rm{^{13}T^{10}_{mb}}} \simeq
\frac{\tau_{21}(\rm{T}_{21}-\rm{T}_{21}^{bg})}{\tau_{10}(\rm{T}_{10}-\rm{T}_{10}^{bg})}.
\]
 
The emission ratios between isotopologue transitions with the same J states do not give unique values for the opacities and the abundance ratios separately, so in order to get values to compare, we assume that the [$^{12}$CO/$^{13}$CO] abundance ratio is  near  40, the values  found toward the centers  of NGC 253 (\citealt{Hen2014}) and IC 342 (\citealt{Mei2001}).  We observe an emission  line ratio $^{13}$CO($2-1$)/$^{13}$CO($1-0$) of 1.1 -- 1.5. The emission ratios are consistent with the single dish data from \citet{Isr2014}.
 
 We discuss five spectral features along the molecular arms that contain detections of $^{12}$CO($2-1$), $^{13}$CO($2-1$), C$^{18}$O($2-1$), and $^{13}$CO($1-0$): {\bf A-1}, {\bf B-1}, {\bf F-2}, {\bf G-2}, and {\bf H-1}.   By starting near an abundance ratio [$^{12}$C/$^{13}$C] = 40, we adjust the abundance ratio and opacity for the isotopologue ratios until a consistent solution is found with the observed intensity ratios.  We find best fit values for the abundance ratios for features {\bf A-1} and {\bf H-1} of [$^{12}$CO/$^{13}$CO] = 50, slightly raised from the value near the center of NGC 253 and IC 342 and [$^{16}$O/$^{18}$O] of about 300.  The [$^{16}$O/$^{18}$O] abundance ratio for the nucleus of NGC 253 was found to be $\sim$ 200 by \citet{Hen1993} which is consistent with the inner Galaxy \citep{Wil1994}.  Opacities for $^{12}$CO(2-1) of around 50 and for $^{13}$CO(2-1) of around unity are implied for both {\bf A-1} and {\bf H-1}.  Spectral feature {\bf F-2} requires the  [$^{16}$O/$^{18}$O] abundance ratio to change from [$^{16}$O/$^{18}$O]=300 to [$^{16}$O/$^{18}$O]=350 to maintain agreement  with the [$^{12}$CO/$^{13}$CO] abundance. For {\bf F-2}, $^{12}$CO opacities of around 15 are implied.  The isotopic ratios for spectral features {\bf B-1} and {\bf G-2} exhibit more extreme values.  To match the ratios with [$^{12}$CO/$^{13}$CO] around 40, [$^{16}$O/$^{18}$O] would be pushed to $>$800, while keeping [$^{16}$O/$^{18}$O] around 300 would require [$^{12}$CO/$^{13}$CO] to drop to $<$20.  Opacities for $^{12}$CO for features {\bf B-1} and {\bf G-2} are around 2 and 12, respectively.  
 
Using the $^{13}$CO($2-1$) to $^{13}$CO($1-0$) ratio, we can constrain the excitation temperature of the gas (\citealt{Aal1995,Mei2001}).  Spectral features {\bf A-1}, {\bf B-1}, {\bf F-2} and {\bf H-1} have   $^{13}$CO  excitation temperatures over 10K, while the feature {\bf G-2} has an excitation temperature of  $\sim$  20 K in the optically thin limit.  For the CND, the limits from the $^{13}$CO($2-1$) to $^{13}$CO($1-0$) ratio give lower limits on the  excitation  temperature of the gas of about 15 K for the outer regions of the CND and 10 K for the inner regions  assuming optically thin gas. 

\subsubsection{Dense Gas in the CND} \label{Sec:DenseGas}

HCN, HCO$^{+}$, HNC, CN and CCH are molecules that have high H$_{2}$ critical densities n$_{crit}$\footnote{The critical density here is listed as the density at which the spontaneous emission coefficient ($A_{ij}$, between levels i and j) is equal to the collision rate [where we adopt a collision coefficient $C_{ij}$ for gas at 10K], n$_{crit} \simeq \frac{A_{ij}}{C_{ij}(10\rm{K})}$.   Radiative trapping and collisions with electrons may reduce the effective critical density of these transitions significantly below the quoted values \citep{Shi2015,Gol2017}. }.  The critical densities of these molecules for their respective transitions are CCH: $1.4 \times 10^{5}$ cm$^{-3}$, HCO$^{+}$: $1.9 \times 10^{5}$ cm$^{-3}$, HCN: $8.7 \times 10^{5}$ cm$^{-3}$, CN: $1.3 \times 10^{6}$ cm$^{-3}$, and HNC: $2.9 \times 10^{6}$ cm$^{-3}$ (\citealt{Gre1974,Flo1999,Liq2010}).   Unlike the  low-J transitions of CO, the J=1-0 transition of  these molecules are excited at gas densities  $\gtrsim 10^{4}$ cm$^{-3}$.  They  dominate  emission from  the CND but not the molecular arms. So the average mass-weighted gas density is higher in the CND than in the molecular arms.

More subtle distinctions between the dense gas tracers can provide insights into the physical and chemical conditions of the CND. Ratios between HCN, HCO$^{+}$, and HNC provide constraints on the gas density and sources of gas chemistry/excitation.  The HCN/HCO$^{+}$, HCN/HNC and HNC/HCO$^{+}$  J=1-0  line ratios tend to be of order unity, with  possible  deviations up to an order of magnitude (e.g. \citealt{Kri2008,Baa2008}).  Various models of photon-dominated regions (PDRs), X-ray dominated regions (XDRs), cosmic ray dissociation regions, and mechanical heating / high temperature chemistry have been  applied  to explain the observed ratios and their  variations (e.g. \citealt{Meij2006, Mei2007, Loe2008, Hara2010, Meij2011}).  CN and CCH are generally thought to trace molecular gas in a partially ionized (mainly C$^{+}$) state associated with PDRs and XDRs (\citealt{Woo1980, Ste1995, Rod1998, Mei2005, Bog2006, Mei2007}).

The behavior of these dense gas line ratios can be subtle and complex, depending on the nature and form of the energetics controlling gas excitation/ionization, but some summarizing statements are possible.  XDRs tend to enhance HCO$^{+}$ modestly and CN strongly relative to HCN and HNC, compared to PDRs, producing HCN/HCO$^{+}$ ratios below unity. However, PDRs with densities of $\sim 10^{4}$ cm$^{-3}$ and high columns can also produce HCN/HCO$^{+} < 1$ (e.g. \citealt{Mei2007}). New calculations of photodissociation rates for HNC find vaues larger than HCN \citep{Agu2017}, indicating PDRs probably favor HCN/HNC $>$ 1, consistent with what is seen in the nearby starburst IC342 \citep{Mei2005}. Increasing cosmic ray ionization generally favors increasing abundances of HCO$^{+}$, relative to HCN, at high densities ($n \sim 10^{5}$ cm$^{-3}$) (e.g. \citealt{Meij2011}).  The addition of mechanical heating or high temperature chemistry tends to elevate HCN and reduce HNC and HCO$^{+}$ abundances (e.g. \citealt{Sch1992, Loe2008, Hara2010, Izu2013}).

Comparing the observed peak values of the dense gas tracers  to theoretical work, excitation sources and conditions can potentially  be constrained. Throughout the map, we find that HCO$^{+}$(1-0) is brighter than HCN(1-0) with an HCN(1-0)/HCO$^{+}$(1-0) ratio near 0.6 everywhere they are both detected. The observed HCN(1-0)/HCO$^{+}$(1-0) ratio is on the low end compared to what is found for Seyfert galaxies such as NGC\,1097 and NGC\,1068 (\citealt{Koh2003}).  \citet{Esp2017} find HCN($4-3$)/HCO${+}$($4-3$) integrated intensity emission ratio is also very low ($<0.5$) in molecular clouds a few tens of pc away from the center in the nuclear filaments, unlike other low luminosity AGNs.   We can place Cen A on Figure 1 of \citet{Koh2003} by adopting an I[$^{12}$CO(2-1)]/I[$^{12}$CO(1-0)] ratio of 0.9.  Using our observed HCN(1-0)/HCO$^{+}$(1-0) ratio and the $^{12}$CO(2-1) from the science verification data, Cen A would be located among the galaxies labeled as ``Composite'' AGNs \citep{Koh2003}.

The HCN(1-0)/HNC(1-0) intensity ratio is typically near unity in molecular environments (e.g. \citealt{Aal2002, Baa2008}). I[HCN(1-0)]/I[HNC(1-0)] ratios toward the CND of Cen A are between 1.5 -- 3.5, strongly favoring HCN over HNC and consistent with densities of $\sim 10^{4}$ cm$^{-3}$ and strong radiation fields / high temperatures.  The combination of the HCN/HCO$^{+}$, HCN/HNC and HCO$^{+}$/HNC ratios also place Cen A  in the PDR parameter space of \citet{Baa2008}, but HCO$^{+}$ remains near the top (XDR) envelope of the PDR range.  This is similar to results by \citet{Koh2003}, including ratios with HNC favor PDRs, but with some "composite" (AGN and starburst) nature.

In the evolved pure starburst M82, HCO$^{+}$ is indeed brighter than HCN \citep{Ngu1992}, so elevated HCO$^{+}$/HCN ratios may not exclusively reflect XDR conditions.   ``Composite'' AGNs containing a nuclear starburst (SB) may enhance HCO$^{+}$ emission through supernova explosions elevating the intense cosmic ionization flux (e.g. \citealt{Ngu1992,Meij2011}).  \citet{Mar2000} detected Pa$\alpha$ in the nuclear region of Cen A and suggests the star formation in the circumnuclear region to be $\sim$ 0.3 M$_{\odot}$ yr$^{-1}$.

Recent results from NGC 1068 \citep{Gar2014} and NGC 1097 \citep{Mar2015} suggest that the differentiation of AGN/SB might be attributable to gas density or mechanical heating \citep{Loe2008,Izu2013,Izu2016} instead of AGN heating.  In fact, adding some mechanical heating to PDRs models \citep{Loe2008} do a better job of matching Cen A's HCN/HCO$^{+}$, HCN/HNC, and HNC/HCO$^{+}$ ratios.

The HCN(1-0)/CN(1-0) ratio is also a useful diagnostic.  Typically CN is used as a tracer of PDRs (\citealt{Gre1996,Rod1998,Bog2006}).  We observe HCN/CN ratios slightly lower in the CND than the arms, with values for the CND of between $\sim$ 2 - 2.5, consistent with expected values that are found toward PDRs.  XDR gas is also predicted to be highly elevated in CN \citep{Meij2006} because XDRs set up PDR-like conditions throughout the bulk of the cloud, not just at its surface. So if XDRs dominated we would expect an even lower HCN/CN ratio.  We conclude that while the observed HCN/CN ratio demonstrates the presence of pronounced photodissociation, it argues in favor of PDRs or soft XDRs over intense XDRs.

CCH is also present in PDR gas and is detected in the CND.  CCH is typically found near high radiation environments (e.g. \citealt{Mei2005,Gar2017}) and although weakly detected in the CND, it could locate a region of high far UV radiation in the southeastern side of the CND.

In summary, we cannot unambiguously determine the nature of the ionization in the CND based on these line ratios, but we can conclude that the CND gas is denser and more strongly irradiated than the molecular arms.  The source of this radiation could be coming from either nuclear star formation / supernovae or from (weak) XDRs associated with the AGN.  The PDR origin is weakly favored over XDRs based on CN and CCH.

\subsubsection{The Influence of the AGN on Gas Chemistry} \label{Sec:AGN_Chem}

The proximity of NGC5128  allows smaller  linear distances  from  the AGN  to be probed  than studies dealing with  the three times more distant  NGC 1068 or NGC 1097 at the same angular resolution.  This allows studies of similar angular resolution to probe closer to the AGN.   The CND has a  couple of  characteristics that could  possibly suggest an interaction with the AGN:  (1)  There is a decreased abundance of the  rare  CO isotopologues in the CND compared to the dense gas tracers  or a high excitation shifting the populations of the rare CO isotopologues to higher rotational transitions \citep{Isr2014}.   (2) The dense gas chemistry suggests that the CND has  PDR/XDR  qualities. 

Bright dense gas tracers in the CND demonstrate that there are large amounts of  dense gas present (See Section \ref{Sec:Masses}).  So, the faintness of the  rare  CO isotopologues in the CND is unlikely to be due to  sub-thermal  excitation.  However, non-LTE effects could contribute to the missing isotopologues.  A possible explanation for this faintness could be isotope-selective photo-dissociation due to the radiation field of the AGN or star formation.  The radiation field could dissociate  much  of the  rare  CO isotopologues throughout the molecular gas in the CND,  but not the $^{12}$CO because it is much more strongly self-shielded.   The fact that the  rare  CO isotopologues are seen in the narrow absorption LV complex gas (but not strongly in the broad HV complex absorption gas) implies that there must remain dense cores along the line of sight that have some associated  rare  CO isotopologues emission.

  Strong radiation likely photodissociates some of the dense gas tracers, particularly HNC and HCN, though less so HCO$^{+}$. (This may be another contributor to the somewhat elevated HCO$^{+}$ line relative to HCN and HNC [Section \ref{Sec:DenseGas}]). But HCN and HCO$^{+}$ can maintain high abundances in PDRs, and have much higher opacities than the rare CO isotopologues (Figure \ref{Spec:abs}) and therefore may partially self-shield, which potentially explains their presence despite the photodissociation.  Anomalously faint $^{13}$CO($1-0$) seems to be a common feature of the nuclei of numerous galaxies (e.g. \citealt{ You1986,Cas1992,Aal1995,Aal1997}).

Another possible, though much more speculative explanation for faint  rare CO isotopologues in the CND, is that the gas in the nucleus of Cen A is the result of accretion of lower metallicity, high CO isotopic ratio gas from outside Cen A.  Lower metallicity gas appears to favor HCO$^{+}$ over HCN because [O/N] increases with decreasing metallicity \citep[e.g.][]{And2014}. However, this does not explain the `normal' galaxy center isotopic ratios implied for the molecular arms, unless the CND and the molecular arms originate in two separate accretion events.  With the current data presented here, we do not see strong evidence for this possibility, but cannot rule it out. 

There is no significant difference in any of the dense gas emission ratios of the inner peaks (features {\bf C-3} and {\bf D-1}) to the outer peaks of the CND ({\bf B-3} and {\bf E-1}).  This indicates that the chemical sphere of influence of the AGN on the molecular gas must span the entirety of the CND or be confined to inside $\lesssim$ 3$"$ currently hidden by the absorption toward the AGN.   The presence of PDR qualities throughout the CND suggests that either the AGN is actively influencing the entire disk, or something else is causing the PDR features. There have been direct observations of influences of an AGN jet on the molecular gas in NGC 1068 \citep{Gar2014} and M51 \citep{Mat2015,Que2016}  so an interaction in Cen A could be expected.  However,  \citet{Mar2000} calculated a maximum star formation rate in the inner kiloparsec of Cen A to be around $1.0 M_{\odot}$ yr$^{-1}$ based on the Pa$\alpha$ emission.  This value includes regions of the molecular arms that are currently forming stars, so the star formation that would interact primarily with CND is less than $1.0 M_{\odot} $yr$^{-1}$.  This star formation could result in increased ionization rates and shocks required to explain the observed line ratios. So, there might be an influence  between the  AGN and the molecular gas in the CND, but we can not rule out star formation or shocks to explain the observed line ratios without a better constraint on the star formation rate in the CND. 

\subsection{Mass Analysis}

\subsubsection{Dynamical Masses} \label{Sec:Masses}

Using the CND as traced by the HCO$^{+}$(1-0) image cube, we estimate the enclosed mass between roughly locations {\bf B} and {\bf E} and locations {\bf C} and {\bf D} using the equation $M_{enc} = \frac{v^2 R}{G \sin^{2}i}$ where $v$ is projected rotational velocity measured at each location, $R$ is the projected distance to that location from the center of the galaxy, $G$ is the gravitational constant, and $i$ is the inclination of the disk.  Spectra were obtained toward these regions and Gaussians were fit to the spectra in order to determine the mean velocity of the gas at these locations.   Locations {\bf B} and {\bf E} trace the outer regions of the CND and probe gas at a radius of 145  $\pm 15 $  pc from the center of the disk, while locations {\bf C} and {\bf D} are the inner peaks of the CND and probe gas at a radius of about 65  $\pm 15 $  pc from the center.  We measured a velocity difference of 355  $\pm 20$  km s$^{-1}$ between {\bf B} and {\bf E} and 185  $\pm 20$  km s$^{-1}$ between {\bf C} and {\bf D}.  We use the reported inclination angle of 71$^{\circ}$ in \citet{Esp2009} and \citet{Qui2010} for the outer region of the CND ({\bf B} and {\bf E}) to determine an enclosed dynamical mass of about  $1.4 \pm 0.2 \times 10^{9} M_{\odot}$ . The uncertainty arises mainly from uncertainties in the inclination angle  of about $\pm 10^{\circ}$.  For the inner portion of the CND ({\bf C} and {\bf D}), we use the reported value of 37$^{\circ}$ for the inclination angle from \citet{Qui2010} to determine an enclosed dynamical mass of  $3.6 \times 10^{8} M_{\odot}$  (uncertain by a factor of 2) within the inner 130 pc of the CND. These values are greater than the  mass of the SMBH  ($5.5\times 10^{7} M_{\odot}$, \citealt{Cap2009}) by factors of  25  for the entire CND and about  6  between the inner peaks of the CND.  Thus the SMBH only begins to dominate the mass budget on scales  much  smaller than R $\sim$ 50 pc.  

\subsubsection{Conversion Factor Analysis}

The column  density of  H$_{2}$,  N$_{\rm{H}_{2}}$, can be calculated by using a conversion factor X  between  the $^{12}$CO($1-0$) intensity and N$_{\rm{H_{2}}}$. We can estimate the  N$_{\rm{H}_{2}}$  from the $^{12}$CO($2-1$) science verification data by assuming a ratio of $^{12}$CO($2-1$) to $^{12}$CO($1-0$) of 0.8 (\citealt{Ler2009}) as both transitions are optically thick. If we adopt the standard Galactic disk conversion factor of $X = 2.0 \times 10^{20}$ cm$^{-2}$ (K km s$^{-1}$)$^{-1}$ (\citealt{Str1988,Hun1997,Bol2003}),  an H$_{2}$ mass of  $1.2\times 10^{8} M_{\odot}$ is calculated for the CND region.  This value was obtained from the integrated intensity map of  the  $^{12}$CO($2-1$) transition  of the inner 500 pc of NGC 5128,  which includes both molecular arm and the CND emission, so this is likely an upper limit on the H$_{2}$ mass in the CND. This H$_{2}$ mass is about  10\%  of the residual mass between the calculated dynamical mass of the CND and M$_{\rm{SMBH}}$.  Using the conversion factor for more energetic environments from \citet{Dow1998} ($X = 0.4 \times 10^{20}$ cm$^{-2}$ (K km s$^{-1}$)$^{-1}$), we calculate the H$_{2}$ mass to be $2.4\times 10^{7} M_{\odot}$, which is about  2\%  of the residual mass between the dynamical mass and the supermassive black hole.  \citet{Isr2014}  use  LVG and PDR models to determine the conversion factor for the CND in NGC 5128 to be $X = 4.0 \times 10^{20}$ cm$^{-2}$ (K km s$^{-1}$)$^{-1}$.  Using this value, we estimate the H$_{2}$ mass to be $2.4\times 10^{8} M_{\odot}$, which is  about 20\% of  the difference between the calculated dynamical and the SMBH  mass.   

The H$_{2}$ column can also be calculated from the $^{13}$CO data if the excitation temperature, opacities, and abundance ratios are known.  The temperature can be determined using the $J = 2-1$ and $J = 1-0$ rotational transitions of $^{13}$CO and the opacities and abundance ratios are determined through isotopologue analysis.  For the CND, where we do not detect $^{13}$CO($1-0$), we get a limit on the column.  If we adopt an excitation temperature for the CND of 20 K (Section  \ref{Sec:Isotope}), and use an abundance ratio of [$^{12}$CO/$^{13}$CO]  $= 50$  and [$^{12}$CO/H$_{2}$] $=8.5 \times 10^{-5}$ (\citealt{Fre1982}), a total gas mass of $\sim 2\times 10^{6} M_{\odot}$ for the CND is calculated.  This value is lower by a factor of $\gtrsim 10$ than the gas masses calculated using $^{12}$CO($2-1$) and {\it any} of the above conversion factors.   This discrepancy could be due to one or several reasons including our adopted [$^{12}$CO/$^{13}$CO] abundance ratio could be incorrect (isotope-selective photo-dissociation may play an important role; see Section \ref{Sec:AGN_Chem}), the line excitation is much different from 20 K, or the conversion factor may be much lower. 

If we consider the molecular arm features individually, we can estimate the H$_{2}$ column toward spectral features {\bf A-1}, {\bf B-1}, {\bf F-2}, {\bf G-2} and {\bf H-1}.  From the isotopologue analysis, the temperatures of the molecular arm features are around 10K for features {\bf A-1}, {\bf B-1}, {\bf F-2}, and {\bf H-1}, but over 20K for {\bf G-2}.  We find that the H$_{2}$ column is also  universally lower  when  calculated from $^{13}$CO($2-1$) than from $^{12}$CO($2-1$) towards the molecular arms.  For the spectral features {\bf A-1}, {\bf F-2}, {\bf G-2} and {\bf H-1}, the calculated values are  only a factor of about 2 lower, which is likely within the uncertainties of the analysis.  However, for region {\bf B-1}  the $^{13}$CO($2-1$) determined H$_{2}$ column is lower than the $^{12}$CO($2-1$) calculated value by around 20 suggesting there are different physical conditions (excitation and opacity) for this feature compared to the rest of the molecular arm components.  

\subsection{Comparing Absorption and Emission}

Our observations include absorption as well as emission.   A detailed  analysis  of the individual absorption components  and their chemisty  will be discussed in a forthcoming paper.  Here  we focus on the line ratios in absorption that directly confront the emission in order to  address  which emission component is most likely responsible for the absorption features.    Absorption depth is converted to line opacity (Section \ref{abssec}), then optical depth ratios are calculated for comparison with the emission line ratios.  As long as temperature/excitation does not change significantly between the two transitions making up the ratios  and assuming the transitions are optically thin,  then the $\tau$ absorption ratio and the T emission ratios may be compared robustly.  Hence if the emission ratios and the absorption ratios are similar this indicates the same gas component is being sampled. 

Ratios of the calculated optical depth $\tau$ were taken from the absorption spectra shown in Figure \ref{Spec:abs} and are reported in Table \ref{tab:absratiodlux}.  Three deep absorption lines in the low velocity complex (LV complex), located at  539.7, 543.4, and 549.7  km s$^{-1}$, and the median of the spectra from velocities 576 -- 604 km s$^{-1}$ to represent the broad, high velocity absorption complex (HV complex) are discussed.   The opacity ratio $\tau$[$^{13}$CO(2-1)]/$\tau$[$^{13}$CO(1-0)] ranges from about 1 to 2 in the absorption and the emission  intensity ratio ranges from 1 to 2 in the molecular arms, but the non-detection of $^{13}$CO(1-0) in the CND puts the opacity ratio at a lower limit of about 2  (emission intensity ratios taken from Table \ref{tab:ratio}).  

The opacity ratio of $\tau$[$^{13}$CO(2-1)]/$\tau$[C$^{18}$O(2-1)] is around 30 in the narrow LV complex components of the absorption and $>$2.5 in the HV complex (Table \ref{tab:absratiodlux}).  In emission, this ratio ranges from about 4-20 in the molecular arms and a lower limit of 4  in the CND (Table \ref{tab:ratio}).  For spectral features {\bf B-1} and {\bf G-2}, the optical depth ratios from the isotopologue analysis would be $\sim$ 20.   Based on this CO isotopologue ratio,   the narrow LV complex  could  be associated with the molecular arms and the HV complex with the CND, but this conclusion remains ambiguous from CO isotopologues alone and runs into problems when considering the dense gas tracers  as discussed below.

The $^{13}$CO(1-0)/HCN(1-0) optical depth ratio is very low in all absorption cases, ranging from 0.04 in the HV complex to about 0.1-0.5 in the narrow LV complex lines.  The emission components from the molecular arms all have an emission ratio greater than about 3.5 for  $^{13}$CO(1-0)/HCN(1-0), while the ratio for the CND is an upper limit around 0.2 for the outer peaks and 0.1 for the inner peaks.  

The single dish data of \citet{Isr2014} has beam sizes of 47$"$ and 57$"$ for $^{13}$CO($1-0$) and HCN($1-0$), respectively.  Assuming emission is uniform over the HCN beam, the ratio is 4.6. If the emission is more compact, this ratio could be as low as 3. This suggests that there is  $\lesssim 25$\% extended emission that is not recovered by our interferometric ALMA observations.  Therefore, we consider the uncertainties of ratios between diffuse CO and the more concentrated high density tracers to be at most 30\%.  Ratios between different CO transitions or between dense gas tracers should be more certain that this.   

There is a strong bimodality between the CND and the molecular arms in the  $^{13}$CO(1-0)/HCN(1-0) emission ratio, with values of less than 0.2 for the CND and values of 3 or larger for the molecular arms.  All of the other ratios seem to be roughly consistent between the two gas regions  (Tables \ref{tab:ratio} and \ref{tab:absratiodlux}).  So based on the $^{13}$CO(1-0)/HCN(1-0) ratio, the ratios of the opacities in the  LV absorption  components  follow more closely the CND values than the corresponding values for the molecular arms, suggesting that the gas in the LV complex absorption is also more likely related to the CND rather than the molecular arms.  However, there is an important caveat, namely absorption is sensitive to molecules in the J=0 state while emission is sensitive to gas in the J=1 state.  Therefore, it is possible that if there is a large amount of very cold/subthermal gas then the HCN absorption will be significantly elevated relative to $^{13}$CO.  If this is the case, then it may still be possible for the absorption components to result from molecular gas farther out than the CND. However, this effect should not significantly affect the comparison of the ratios between the dense gas tracers. 

The absorption component centered on roughly 543 km s$^{-1}$ seems to be different from the other two narrow LV complex components in a number of ratios.  This component  has a  much narrower line width than the other two and is the only region in both absorption  or  emission, where  HCN(1-0) is stronger than HCO$^{+}$(1-0).   The differences in the ratios of the other dense gas tracers for this component suggest that HCN(1-0) is enhanced compared to the other narrow LV complex components.   The pencil beam for the absorption could be picking out a specific region of a molecular cloud that happens to deviate from the average that is observed in emission.  

 The optical depth ratio of HCN(1-0)/CN(1-0) toward the narrow LV components are 1.1 -- 1.7, which is similar to the CND emission values of 1.7 ({\bf C-3}).  Likewise for the HCN(1-0)/CCH(1-0) opacity ratios, the narrow LV components are 3.0 -- 6.6, consistent with the emission value of 3.9.  Therefore, these ratios also suggest that the narrow line LV components could be associated with the CND.  The HV component is further  enhanced in CN and CCH relative to the CND/LV (HV: HCN(1-0)/CN(1-0) $\simeq$ 0.7 and HCN(1-0)/CCH(1-0) $\simeq$ 1.5).  CN and CCH are both molecules that are present  in PDR/XDRs, which indicates that the HV complex component is located in a region of stronger radiation field, possibly closer to the AGN.  

Combining this with the previous statement that the gas seen in absorption seems to be similar to the gas that is observed in the CND,  suggest s  the absorption features  may  be placed entirely in the CND.  The narrow LV complex components, which are at velocities much closer to systemic,  must  be coming from gas crossing in front of the AGN  with small noncircular motions.  The redshifted HV complex component of absorption then would come from gas much closer to the AGN.  In fact, \citet{Esp2010} used VLBA observations of H\,{\sc i} to determine that the broad component is very close to the AGN by finding a significant change in the HV complex on scales of 0.4 pc.  \citet{Mul2009} do not observe HCO$^{+}$($4-3$) or HCN($4-3$) in the HV complex and constrain the density of the LV absorption gas to be a few 10$^4$ cm$^{-3}$, consistent with the gas chemistry measured in the CND. 

\subsection{The Warped Disk Model}

\citet{Qui2010} have  compiled  a warped disk model of NGC 5128 that ranges from 2 pc up to 6500 pc using 2.122 $\mu$m molecular hydrogen line  data (inner 3$"$;  \citealt{Neu2007}), SMA CO observations (inner 1$'$; \citealt{Esp2009}), Spitzer 8 $\mu$m ($>$ 40$"$; \citealt{Qui2006}) and H\,{\sc i} observations ($>$ 60$"$; \citealt{Str2010}).  We do not detect the entirety of the  dust lane so we do not  compare our data  to the large scale portions of the model. We do detect the entire CND, which is about 22$"$ $\times$ 11$"$ in size at a position angle of 155$^{\circ}$ and an inclination of about 60$^{\circ}$ for the outer components compared to the position angle of 150$^{\circ}$ and inclination of $\sim$ 70$^{\circ}$ reported in \citet{Qui2010}.  The inner components of the CND are at a radius of about 65 pc from the AGN. 

Using the warped disk model from \citet{Qui2010}, we develop a  potential picture of the inner region of Cen A.  The inner 200 pc of NGC 5128 contains the CND, which is abundant in dense gas.  Beyond the CND, there appears to be a gap in the molecular gas, before reaching the warped rings that create the parallel  molecular  arms, seen prominently in CO and its  rare  isotopologues.  These warps create a ridge of overlapping  molecular gas in the south-eastern and  north-western regions explaining  the molecular arms.   The front and back sides of  the turnover in  the warp could  possibly  explain  the splitting along the arms seen in the PV diagrams  (See Figure \ref{Fig:PVdiag} Slice 3). The regions where fewer of the molecular rings overlap are too faint to be detected by our observations.

The jet from the AGN is roughly perpendicular to the outer part of the CND and we assume that the northern jet is approaching and is brighter due to  Doppler boosting  \citep{Jon1996,Tin1998}  (See also  Figure \ref{Fig:Cont}).  This would mean that the CND, which is at a position angle of roughly 150$^{\circ}$ and inclined at about 60$^{\circ}$, has the  northeastern  side of the disk oriented away from Earth and the southwestern side closer to Earth.  This would put the redshifted side of the CND (the  northwestern  side), which have velocities of $\geq$580 km s$^{-1}$, behind the AGN meaning that the HV complex component in the absorption can not be coming from the redshifted side of the CND, assuming the CND is purely flat with only circular motions.  However, from comparing the chemistry in the absorption to the emission components, the  LV and HV  absorption components appear more similar to the CND than the molecular arms.  

Based on our data, we present two different scenarios that could result in the observed absorption and emission.  In the first scenario, all of the narrow LV complex components of the absorption come from where individual clouds in the CND cross the AGN.  The broad component would then come from material closer to the middle of the CND that is falling in towards the AGN,  that may ultimately  fuel the jet.  If the gas is falling in toward the black hole, one would expect that the gas would be stretched in velocity space as the gas nearer to the black hole is moving at a higher velocity than gas that is further away.  This could  be consistent with  the HV complex  being  similar to one of the narrow components but stretched and redshifted over a range of 25 -- 60 km s$^{-1}$ as the gas falls in toward the black hole.  \citet{Eck1999} described a scenario in which both of the LV and HV complexes originate from gas farther away from the AGN than the CND, however, based on the chemistry of the absorption components, the gas in absorption appears to be more closely related to the CND than gas further away.  

In the second scenario, the warps in the molecular gas of Cen A continue into the innermost part of the CND and it is possible that the warps  are severe enough to  place part of the redshifted side of the CND in front of the AGN to create the broad HV complex component, while some other non-warped parts of the CND remain in front to make the narrow roughly systemic LV complex components. 

The velocities of the inner peaks of the CND are at about 470 and 660 km s$^{-1}$ and at a radius of about 65 pc, which means that the rotational velocity of that tilted ring is 1/2 the difference of those velocities, around 95 km s$^{-1}$.  The maximum redshifted velocity that we observe in the absorption is about 60 km s$^{-1}$ off the systemic velocity, meaning that gas interior to the inner peaks of the CND  would  be responsible for parts of the HV complex component.  However, this gas would need an extreme warp for these high velocities, so non-circular orbits are likely present.  In fact, large non-circular motions are found in the CND by \citet{Esp2017}.    The chemistry of the HV complex component is enhanced in the PDR/XDR tracers CN and CCH.  This would be expected of gas that is getting closer to the radiation field from the AGN.  This is also consistent with the fact that we detect CN emission almost exclusively from the inner CND.

\section{Summary}

We observed various molecular transitions toward the inner 500 pc of Centaurus A  with ALMA and obtained maps and spectra at much higher sensitivity than previous observations.   Based on these observations and analysis performed, we  draw  the following conclusions about the nuclear region of Cen A: 

\begin{itemize}

\item{ {\bf The molecular arms are  kinematically and chemically distinct from the CND.}  The molecular arms are strongly detected in CO and its isotopologues but dense gas tracers like HCO$^{+}$ and HCN were not detected or detected at only very weak levels.  The CND, however, is only weakly detected in the  rare  CO isotopologues, but it is strongly detected in the dense gas tracers.   There is no apparent connection between the emission components corresponding to the molecular arms and the CND.  However, the lack of short baselines and resulting lack of sensitivity to large scale structure may have meant a connection between the components was not detected. Without a clear transition between the molecular arms and the CND, the mechanism for the transportation of molecular gas from the outer galaxy to the nuclear is still missing. 
}

\item{ {\bf Both the high and low velocity complexes in the absorption profiles  appear to originate  from gas in the CND.}   The optical depth ratios from the absorption in both the high and low velocity complexes match the line ratios from the CND and not the  dust lane.  Also, the dense gas tracers that  are  prominent in the CND, but weak or not detected in the molecular arms, were also prominent in absorption.  Placing all of the absorption profiles within the CND means that the velocities from the absorption profiles must be  accounted for by  the  kinematics  of the CND.  
}
\item{ {\bf The high velocity complex of the absorption spectra comes from gas that may be infalling toward the AGN.}  With all of the absorption components placed inside the CND, the high velocity offset of the high velocity complex could be the result of the gas falling in toward the AGN.   If this gas is indeed infalling toward the AGN, then it could potentially be the gas that  will  fuel the central engine that is powering the radio jet.  The HV complex has long been considered as potentially infalling gas based on prior observations of the absorption (e.g. \citealt{Gar1976,Hul1983,Sea1990,Isr1991}). 
}
\item{ {\bf The AGN may be radiatively influencing the CND.}  The lack of  emission in low $J$ level transitions of the  rare  CO isotopologues and bright CN and CCH in the inner CND  suggest that energy is being injected into the CND.  The  poorly constrained  level of star formation in the CND  suggests the  AGN as  a possible  source of the radiation.  If the AGN is indeed responsible for the energy input to the CND, this would be direct  chemical  feedback from the AGN to the molecular gas in the nuclear region of the galaxy. 
}

\end{itemize}

\acknowledgments
We thank the referee for comments that improved the manuscript.  We would like to thank Fabian Walter for stimulating conversations and comments.  D. E. was supported by JSPS KAKENHI Grant Number JP 17K14254.  This manuscript makes use of the following ALMA data: ADS/JAO.ALMA
\#2011.0.00008.SV and 2011.0.00010.S. The National Radio Astronomy Observatory is a facility of the National Science Foundation operated under cooperative agreement by Associated Universities, Inc. ALMA is a partnership of ESO (representing its member states), NSF (USA), and NINS (Japan), together with NRC (Canada), NSC and ASIAA (Taiwan), and KASI (Republic of Korea), in cooperation with the Republic of Chile.  The Joint ALMA Observatory is operated by ESO, AUI/NRAO and NAOJ. M. M. and D. S. M. acknowledges support from National Science Foundation Grant AST-1109803. Mark McCoy is a student at the National Radio Astronomy Observatory.

%
\bibliographystyle{plainnat}

\clearpage

\clearpage

\clearpage

\clearpage

\clearpage 

\clearpage

\clearpage

\clearpage

\clearpage

\clearpage

\clearpage


\begin{deluxetable}{lcccccccc}
\tabletypesize{\scriptsize}
\tablecaption{ Observational Parameters }
\rotate
\tablecolumns{9}
\tablewidth{0pt}
\tablehead{
\colhead{Center Freq.} & \colhead{Bandwidth} & \colhead{$\Delta v$} & \colhead{Date} &
\colhead{Beam} & \colhead{Channel RMS} &   
\colhead{Cont. Flux Density} & \colhead{Spectral} & \colhead{Rest Freq.}
}

\startdata
(GHz) & (MHz) [km s$^{-1}$] & (km s$^{-1}$) &  & ($'' \times ''$,$^{o}$ ) & (mJy beam$^{-1}$) & (Jy) & Line ID & (GHz) \\

(1) & (2) & (3) & (4) & (5) & (6) & (7) & (8) & (9) \\

\sidehead{{\bf Band 3 (3 mm)}} 
85.338706  & 234.4  [806] & 0.420 & 2012 Apr 07  & 2.07 $\times$ 1.69, $-$76.2 & 4.1  & 7.5 &  &  \\
86.846824  & 234.4  [822]  & 0.428 & 2012 Apr 07  & 2.19 $\times$ 1.67, $-$83.7 & 4.8  & 7.4 &  &  \\
87.169498  & 234.4  [806]  & 0.420 & 2012 May 08  & 2.19 $\times$ 1.61, 72.2    & 4.8  & 7.8 & {\bf CCH(N$=1-0$;J$=\frac{3}{2}-\frac{1}{2}$)     }       & 87.316925 \\
87.765057  & 234.4  [799]  & 0.416 & 2012 May 08  & 2.17 $\times$ 1.60, 73.8    & 4.2  & 7.9 & HNCO($4_{04}-3_{03}$)           & 87.925238 \\
88.470137  & 234.4  [791]  & 0.412 & 2012 May 08  & 2.13 $\times$ 1.59, 74.6    & 4.0  & 7.4 & {\bf HCN($1-0$)          }  & 88.631847 \\
89.026047  & 234.4 [787] & 0.410 & 2012 May 08  & 2.12 $\times$ 1.57, 74.8    & 4.2  & 7.9  & {\bf HCO$^{+}$($1-0$) }     & 89.188526 \\
90.498407  & 234.4  [776]  & 0.404 & 2012 May 08  & 1.66 $\times$ 1.61, $-$59.0 & 4.4  & 8.1  & {\bf HNC($1-0$)      }      & 90.663564 \\
90.813277  & 234.4  [772]  & 0.402 & 2012 May 08  & 1.66 $\times$ 1.61, $-$46.1 & 3.9  & 8.1 & HC$_{3}$N($10-9$)     & 90.978989 \\
91.803617  & 234.4  [764] & 0.398 & 2012 May 08  & 1.66 $\times$ 1.58, $-$56.6 & 4.0  & 8.2  & CH$_{3}$CN($5_{3}-4_{3}$)     & 91.971160 \\
93.003656  & 234.4  [753]  & 0.392 & 2012 May 08  & 1.63 $\times$ 1.55, $-$73.0 & 3.8  & 8.2 & N$_{2}$H$^{+}$($1-0$) & 93.173777 \\
97.979577  & 234.4  [714] & 0.372 & 2012 Apr 07  & 1.68 $\times$ 1.45, $-$59.4 & 4.6  & 7.6 &  &  \\
100.076592 & 234.4  [703]  & 0.366 & 2012 Apr 07  & 1.88 $\times$ 1.50, $-$70.9 & 5.1  & 7.6 &  &  \\
109.590043 & 234.4 [641] & 0.334 & 2012 May 07  & 1.38 $\times$ 1.35, $-$73.7 & 5.3  & 9.0 & C$^{18}$O($1-0$)      & 109.782176 \\
110.001195 & 234.4  [637] & 0.332 & 2012 May 07  & 1.41 $\times$ 1.32, $-$72.9 & 4.6  & 9.0 & {\bf $^{13}$CO($1-0$)  }    & 110.201354 \\
112.154164 & 234.4  [626] & 0.326 & 2012 May 07  & 1.38 $\times$ 1.30, $-$54.1 & 5.4  & 9.1 & C$^{17}$O($1-0$)      & 112.358988 \\
113.230037 & 234.4  [618] & 0.322 & 2012 May 07  & 1.38 $\times$ 1.27, $-$69.7 & 6.1  & 9.2  & {\bf CN(N$=1-0$;J$=\frac{3}{2}-\frac{1}{2}$)      }       & 113.490982 \\
\sidehead{{\bf Band 6 (1 mm)}}  
218.077608 & 937.5  [1286] & 0.670 & 2012 Jan 24    & 2.22 $\times$ 1.43, $-$44.4 & 4.0  & 6.5& H$_{2}$CO($3_{03}-2_{02}$)      & 218.222195 \\
219.152815 & 937.5  [1279] & 0.666 & 2012 Jan 24    & 2.22 $\times$ 1.43, $-$44.4 & 4.3  & 6.5 & {\bf C$^{18}$O($2-1$)    }  & 219.560358 \\ 
219.397838 & 937.5  [1279] & 0.666 & 2012 Jan 24    & 2.23 $\times$ 1.45, $-$46.0 & 4.7  & 6.5 & HNCO($10_{010}-9_{09}$)          & 219.798282 \\
220.007315 & 937.5  [1275] & 0.664 & 2012 Jan 24    & 2.23 $\times$ 1.46, $-$44.6 & 3.9  & 6.8   & {\bf $^{13}$CO($2-1$) }     & 220.398684 \\
\enddata
\label{tab:obs}
\tablecomments{(1) the center frequencies of the bands. (2)  the bandwidths.  (3) the velocity width of the channels.  (4) the date of the observations. (5) the synthesized beams from imaging with natural weighting.  (6) rms of emission and absorption free  channels  of the image cubes.  (7) the flux densities of the central continuum source.   Overall flux density uncertainties are taken to be 5\% in Band 3 and 10\% in Band 6.  (8) the molecular line  targeted  in emission in the frequency ranges.  (9)  The rest frequencies of the  targeted  molecular transitions from \protect \citet{Lov2004}. Transitions detected in emission are in bold.  For transitions that have  resolved  hyperfine structure, the strongest of the hyperfine is reported.  }

\end{deluxetable}

\begin{deluxetable}{lccccc}
\tabletypesize{\scriptsize}
\tablecaption{ Properties of Emission Components }
\tablecolumns{6}
\tablewidth{0pt}
\tablehead{
\colhead{Line Identifier} & \colhead{Transition} & \colhead{Peak} & \colhead{Line Center} &
\colhead{FWHM } & \colhead{Integral}
}
\startdata

  &  & {\bf (K)} & {\bf (km s$^{-1}$)} & {\bf (km s$^{-1}$)} & {\bf (K km s$^{-1}$)} \\
   & & & & & \\
A-1	&	$^{12}$CO(2-1)	&	$ 0.92 \pm 0.12 $	&	$ 510.0 \pm 5.0 $	&	$ 20.0 \pm 5.0 $	&	$ 19.4 \pm 5.6 $	\\
(13:25:27.983)	&	$^{13}$CO(2-1)	&	$ 0.61 \pm 0.03 $	&	$ 508.5 \pm 0.3 $	&	$ 10.4 \pm 0.6 $	&	$ 6.7 \pm 0.5 $  	\\
(-43.01.01.519)	&	$^{13}$CO(1-0)	&	$ 0.42 \pm 0.03 $	&	$ 513.1 \pm 0.3 $	&	$ 7.8 \pm 0.8 $  	&	$ 3.5 \pm 0.4 $  	\\
	&	C$^{18}$O(2-1)	&	$ 0.14 \pm 0.02 $	&	$ 509.6 \pm 0.6 $	&	$ 8.5 \pm 1.4 $ 		&	$ 1.3 \pm 0.3 $  	\\
	&	C$^{18}$O(1-0)	&	$<0.03$	&	 \nodata	&	 \nodata	&	 \nodata	\\
	&	HCO$^{+}$(1-0)	&	$<0.06$	&	 \nodata	&	 \nodata	&	 \nodata	\\
	&	HCN(1-0)			&	$<0.06$	&	 \nodata	&	 \nodata	&	 \nodata	\\
	&	CN(1-0)			&	$<0.05$	&	 \nodata	&	 \nodata	&	 \nodata	\\
	&	HNC(1-0)			&	$<0.06$	&	 \nodata	&	 \nodata	&	 \nodata	\\
	&	CCH(1-0)			&	$<0.06$	&	 \nodata	&	 \nodata	&	 \nodata	\\

A-2	&	$^{12}$CO(2-1)	&	$ 0.31 \pm 0.05 $	&	$ 550.0 \pm 5.0 $	&	$ 15.0 \pm 5.0 $	&	$ 4.9 \pm 3.6 $ 	\\
	&	$^{13}$CO(2-1)	&	$ 0.26 \pm 0.07 $	&	$ 553.7 \pm 1.3 $	&	$ 10.8 \pm 3.2 $	&	$ 3.0 \pm 1.2 $	\\
	&	$^{13}$CO(1-0)	&	$ 0.18 \pm 0.01 $	&	$ 560.1 \pm 0.4 $	&	$ 4.7 \pm 0.9 $  	&	$ 0.9 \pm 0.2 $	\\
	&	C$^{18}$O(2-1)	&	$ <0.02 $	&	 \nodata	&	 \nodata	&	 \nodata	\\
	&	C$^{18}$O(1-0)	&	$ <0.03 $	&	 \nodata	&	 \nodata	&	 \nodata	\\
	&	HCO$^{+}$(1-0)	&	$ <0.06 $	&	 \nodata	&	 \nodata	&	 \nodata	\\
	&	HCN(1-0)			&	$ <0.06 $	&	 \nodata	&	 \nodata	&	 \nodata	\\
	&	CN(1-0)			&	$ <0.05 $	&	 \nodata	&	 \nodata	&	 \nodata	\\
	&	HNC(1-0)			&	$ <0.06 $	&	 \nodata	&	 \nodata	&	 \nodata	\\
	&	CCH(1-0)			&	$ <0.06 $	&	 \nodata	&	 \nodata	&	 \nodata	\\

A-3	&	$^{12}$CO(2-1)	&	$ 0.15 \pm 0.03 $	&	$ 614.1 \pm 6.3 $ 	&	$61.0 \pm 15.0 $ 	&	$10.0 \pm 3.3 $	\\
	&	$^{13}$CO(2-1)	&	$ <0.02 $	&	 \nodata	&	 \nodata	&	 \nodata	\\
	&	$^{13}$CO(1-0)	&	$ 0.02 \pm 0.01 $	&	$ 631.8 \pm 12.0$ 	&	$40.0 \pm 31.0$	&	$ 0.9 \pm 0.8 $	\\
	&	C$^{18}$O(2-1)	&	$<0.02 $	&	 \nodata	&	 \nodata	&	 \nodata	\\
	&	C$^{18}$O(1-0)	&	$<0.03$	&	 \nodata	&	 \nodata	&	 \nodata	\\
	&	HCO$^{+}$(1-0)	&	$ 0.07 \pm 0.03 $	&	$ 627.1 \pm 3.8 $	&	$ 19.6 \pm 9.0 $	&	$ 1.5 \pm 0.9 $	\\
	&	HCN(1-0)			&	$ <0.06 $	&	 \nodata	&	 \nodata	&	 \nodata	\\
	&	CN(1-0)			&	$ <0.05 $	&	 \nodata	&	 \nodata	&	 \nodata	\\
	&	HNC(1-0)			&	$ <0.06 $	&	 \nodata	&	 \nodata	&	 \nodata	\\
	&	CCH(1-0)			&	$ <0.06$	&	 \nodata	&	 \nodata	&	 \nodata	\\

B-1	&	$^{12}$CO(2-1)	&	$ 2.17 \pm 0.05 $	&	$ 525.9 \pm 0.2 $	&	$ 22.8 \pm 0.6 $	&	$ 52.5 \pm 1.7 $	\\
(13:25:27.490)	&	$^{13}$CO(2-1)	&	$ 0.42 \pm 0.02 $	&	$ 533.1 \pm 0.4 $	&	$ 16.0 \pm 0.9 $	&	$ 7.2 \pm 0.5 $	\\
(-43.01.00.370)	&	$^{13}$CO(1-0)	&	$ 0.29 \pm 0.04 $	&	$ 537.6 \pm 0.9 $	&	$ 13.8 \pm 2.0 $	&	$ 4.3 \pm 0.8 $	\\
	&	C$^{18}$O(2-1)	&	$ 0.02 \pm 0.01 $	&	$ 535.5 \pm 2.2 $	&	$ 13.5 \pm 5.1 $	&	$ 0.5 \pm 0.2 $	\\
	&	C$^{18}$O(1-0)	&	$ <0.04 $	&	 \nodata	&	 \nodata	&	 \nodata	\\
	&	HCO$^{+}$(1-0)	&	$ 0.14 \pm 0.03 $	&	$ 534.4 \pm 1.3 $	&	$ 11.1 \pm 3.1 $	&	$ 1.7 \pm 0.6 $	\\
	&	HCN(1-0)			&	$ 0.08 \pm 0.03 $	&	$ 532.7 \pm 2.3 $	&	$ 12.9 \pm 5.5 $	&	$ 1.1 \pm 0.6 $	\\
	&	CN(1-0)			&	$<0.05$	&	 \nodata	&	 \nodata	&	 \nodata	\\
	&	HNC(1-0)			&	$ <0.06 $	&	 \nodata	&	 \nodata	&	 \nodata	\\
	&	CCH(1-0)			&	$ <0.06$	&	 \nodata	&	 \nodata	&	 \nodata	\\

B-2	&	$^{12}$CO(2-1)	&	$ 0.26 \pm 0.05 $	&	$ 570.0 \pm 5.0 $	&	$ 15.0 \pm 5.0 $	&	$ 4.1 \pm 3.6 $	\\
	&	$^{13}$CO(2-1)	&	$ 0.12 \pm 0.03 $	&	$ 573.2 \pm 1.0 $	&	$ 9.5 \pm 2.4 $	         &	$ 1.2 \pm 0.4 $	\\
	&	$^{13}$CO(1-0)	&	$ <0.04$	&	 \nodata	&	 \nodata	&	 \nodata	\\
	&	C$^{18}$O(2-1)	&	$ <0.01$	&	 \nodata	&	 \nodata	&	 \nodata	\\
	&	C$^{18}$O(1-0)	&	$<0.04$	&	 \nodata	&	 \nodata	&	 \nodata	\\
	&	HCO$^{+}$(1-0)	&	$<0.06$	&	 \nodata	&	 \nodata	&	 \nodata	\\
	&	HCN(1-0)			&	$<0.12$	&	 \nodata	&	 \nodata	&	 \nodata	\\
	&	CN(1-0)			&	$<0.05$	&	 \nodata	&	 \nodata	&	 \nodata	\\
	&	HNC(1-0)			&	$<0.06$	&	 \nodata	&	 \nodata	&	 \nodata	\\
	&	CCH(1-0)			&	$<0.06$	&	 \nodata	&	 \nodata	&	 \nodata	\\

{\bf B-3	}&	$^{12}$CO(2-1)	&	$ 1.47 \pm 0.05 $	&	$ 711.7 \pm 1.2 $	&	$ 64.4 \pm 2.9 $	&	$ 100.6 \pm 5.9 $	\\
	&	$^{13}$CO(2-1)	&	$ 0.07 \pm 0.01 $	&	$ 715.4 \pm 2.4 $	&	$ 68.1 \pm 5.6 $	&	$ 5.3 \pm 0.6 $	\\
	&	$^{13}$CO(1-0)	&	$<0.04$	&	 \nodata	&	 \nodata	&	 \nodata	\\
	&	C$^{18}$O(2-1)	&	$<0.01$	&	 \nodata	&	 \nodata	&	 \nodata	\\
	&	C$^{18}$O(1-0)	&	$<0.04$	&	 \nodata	&	 \nodata	&	 \nodata	\\
	&	HCO$^{+}$(1-0)	&	$ 0.40 \pm 0.02 $	&	$ 712.7 \pm 1.5 $	&	$ 65.5 \pm 3.5 $	&	$ 28.3 \pm 2.0 $	\\
	&	HCN(1-0)			&	$ 0.19 \pm 0.02 $	&	$ 701.9 \pm 2.8 $	&	$ 53.8 \pm 6.9 $	&	$ 11.7 \pm 1.8 $	\\
	&	CN(1-0)			&	$ 0.04 \pm 0.01 $	&	$ 697.6 \pm 6.5 $	&	$ 48.0 \pm 15.0 $	&	$ 2.2 \pm 0.9 $	\\
	&	HNC(1-0)			&	$ 0.11 \pm 0.02 $	&	$ 720.0 \pm 5.2 $	&	$ 63.0 \pm 12.0 $	&	$ 7.3 \pm 1.9 $	\\
	&	CCH(1-0)			&	$ 0.07 \pm 0.02 $	&	$ 726.9 \pm 5.7 $	&	$ 36.0 \pm 13.0 $	&	$ 2.7 \pm 1.3 $	\\

C-1	&	$^{12}$CO(2-1)	&	$<0.12$	&	 \nodata	&	 \nodata	&	 \nodata	\\
(13:25:27.379)	&	$^{13}$CO(2-1)	&	$ 0.03 \pm 0.02 $	&	$ 523.6 \pm 3.5 $	&	$ 10.8 \pm 8.2 $	&	$ 0.3 \pm 0.3 $	\\
(-43.01.05.728)	&	$^{13}$CO(1-0)	&	$<0.04$	&	 \nodata	&	 \nodata	&	 \nodata	\\
	&	C$^{18}$O(2-1)	&	$<0.01$	&	 \nodata	&	 \nodata	&	 \nodata	\\
	&	C$^{18}$O(1-0)	&	$<0.04$	&	 \nodata	&	 \nodata	&	 \nodata	\\
	&	HCO$^{+}$(1-0)	&	$<0.06$	&	 \nodata	&	 \nodata	&	 \nodata	\\
	&	HCN(1-0)			&	$<0.06$	&	 \nodata	&	 \nodata	&	 \nodata	\\
	&	CN(1-0)			&	$<0.04$	&	 \nodata	&	 \nodata	&	 \nodata	\\
	&	HNC(1-0)			&	$<0.06$	&	 \nodata	&	 \nodata	&	 \nodata	\\
	&	CCH(1-0)			&	$<0.06$	&	 \nodata	&	 \nodata	&	 \nodata	\\

C-2	&	$^{12}$CO(2-1)	&	$ 0.19 \pm 0.05 $	&	$ 560.0 \pm 5.0 $	&	$ 15.0 \pm 5.0 $	&	$ 3.1 \pm 3.6 $	\\
	&	$^{13}$CO(2-1)	&	$ 0.18 \pm 0.05 $	&	$ 557.5 \pm 5.0 $	&	$ 5.0 \pm 3.0 $	&	$ 0.9 \pm 0.8 $	\\
	&	$^{13}$CO(1-0)	&	$<0.04$	&	 \nodata	&	 \nodata	&	 \nodata	\\
	&	C$^{18}$O(2-1)	&	$<0.01$	&	 \nodata	&	 \nodata	&	 \nodata	\\
	&	C$^{18}$O(1-0)	&	$<0.04$	&	 \nodata	&	 \nodata	&	 \nodata	\\
	&	HCO$^{+}$(1-0)	&	$<0.06$	&	 \nodata	&	 \nodata	&	 \nodata	\\
	&	HCN(1-0)			&	$<0.06$	&	 \nodata	&	 \nodata	&	 \nodata	\\
	&	CN(1-0)			&	$<0.04$	&	 \nodata	&	 \nodata	&	 \nodata	\\
	&	HNC(1-0)			&	$<0.06$	&	 \nodata	&	 \nodata	&	 \nodata	\\
	&	CCH(1-0)			&	$<0.06$	&	 \nodata	&	 \nodata	&	 \nodata	\\

{\bf C-3	}&	$^{12}$CO(2-1)	&	$ 1.04 \pm 0.09 $	&	$ 649.1 \pm 1.9 $	&	$ 46.4 \pm 4.5 $	&	$ 51.2 \pm 6.6 $	\\
	&	$^{13}$CO(2-1)	&	$ 0.04 \pm 0.01 $	&	$ 646.7 \pm 2.4 $	&	$ 46.4 \pm 5.6 $	&	$ 2.1 \pm 0.3 $	\\
	&	$^{13}$CO(1-0)	&	$<0.04$	&	 \nodata	&	 \nodata	&	 \nodata	\\
	&	C$^{18}$O(2-1)	&	$<0.01$	&	 \nodata	&	 \nodata	&	 \nodata	\\
	&	C$^{18}$O(1-0)	&	$<0.04$	&	 \nodata	&	 \nodata	&	 \nodata	\\
	&	HCO$^{+}$(1-0)	&	$ 0.64 \pm 0.02 $	&	$ 651.7 \pm 1.0 $	&	$ 56.3 \pm 2.2 $	&	$ 38.4 \pm 2.0 $	\\
	&	HCN(1-0)			&	$ 0.37 \pm 0.02 $	&	$ 648.4 \pm 1.4 $	&	$ 50.7 \pm 3.4 $	&	$ 20.0 \pm 1.8 $	\\
	&	CN(1-0)			&	$ 0.21 \pm 0.01 $	&	$ 649.2 \pm 1.4 $	&	$ 42.3 \pm 3.4 $	&	$ 9.6 \pm 1.0 $	\\
	&	HNC(1-0)			&	$ 0.19 \pm 0.02 $	&	$ 650.2 \pm 2.4 $	&	$ 38.0 \pm 5.8 $	&	$ 7.6 \pm 1.5 $	\\
	&	CCH(1-0)			&	$ 0.10 \pm 0.02 $	&	$ 664.8 \pm 5.7 $	&	$ 64.0 \pm 14.0 $	&	$ 6.6 \pm 1.8 $	\\

{\bf D-1 }	&	$^{12}$CO(2-1)	&	$ 0.86 \pm 0.05 $	&	$ 468.4 \pm 1.2 $	&	$ 42.0 \pm 2.9 $	&	$ 38.4 \pm 3.6 $	\\
(13:25:27.918)	&	$^{13}$CO(2-1)	&	$ 0.04 \pm 0.01 $	&	$ 459.1 \pm 1.8 $	&	$ 35.4 \pm 4.3 $	&	$ 1.4 \pm 0.2 $	\\
(-43.01.09.025)	&	$^{13}$CO(1-0)	&	$<0.03$	&	 \nodata	&	 \nodata	&	 \nodata	\\
	&	C$^{18}$O(2-1)	&	$<0.01$	&	 \nodata	&	 \nodata	&	 \nodata	\\
	&	C$^{18}$O(1-0)	&	$<0.04$	&	 \nodata	&	 \nodata	&	 \nodata	\\
	&	HCO$^{+}$(1-0)	&	$ 0.83 \pm 0.02 $	&	$ 466.9 \pm 0.7 $	&	$ 46.3 \pm 1.6 $	&	$ 40.8 \pm 1.8 $	\\
	&	HCN(1-0)			&	$ 0.48 \pm 0.02 $	&	$ 465.2 \pm 1.2 $	&	$ 49.7 \pm 2.9 $	&	$ 25.5 \pm 2.0 $	\\
	&	CN(1-0)			&	$ 0.20 \pm 0.02 $	&	$ 465.9 \pm 1.7 $	&	$ 27.7 \pm 4.1 $	&	$ 5.8 \pm 1.2 $	\\
	&	HNC(1-0)			&	$ 0.16 \pm 0.02 $	&	$ 458.8 \pm 1.8 $	&	$ 39.6 \pm 4.3 $	&	$ 6.9 \pm 1.0 $	\\
	&	CCH(1-0)			&	$<0.07$	&	 \nodata	&	 \nodata	&	 \nodata	\\

D-2	&	$^{12}$CO(2-1)	&	$ 0.36 \pm 0.05 $	&	$ 530.0 \pm 5.0 $	&	$ 20.0 \pm 5.0 $	&	$ 7.7 \pm 4.6 $	\\
	&	$^{13}$CO(2-1)	&	$<0.01$	&	 \nodata	&	 \nodata	&	 \nodata	\\
	&	$^{13}$CO(1-0)	&	$<0.03$	&	 \nodata	&	 \nodata	&	 \nodata	\\
	&	C$^{18}$O(2-1)	&	$<0.01$	&	 \nodata	&	 \nodata	&	 \nodata	\\
	&	C$^{18}$O(1-0)	&	$<0.04$	&	 \nodata	&	 \nodata	&	 \nodata	\\
	&	HCO$^{+}$(1-0)	&	$<0.07$	&	 \nodata	&	 \nodata	&	 \nodata	\\
	&	HCN(1-0)			&	$<0.06$	&	 \nodata	&	 \nodata	&	 \nodata	\\
	&	CN(1-0)			&	$<0.06$	&	 \nodata	&	 \nodata	&	 \nodata	\\
	&	HNC(1-0)			&	$<0.06$	&	 \nodata	&	 \nodata	&	 \nodata	\\
	&	CCH(1-0)			&	$<0.07$	&	 \nodata	&	 \nodata	&	 \nodata	\\

D-3	&	$^{12}$CO(2-1)	&	$<0.23$	&	 \nodata	&	 \nodata	&	 \nodata	\\
	&	$^{13}$CO(2-1)	&	$ 0.02 \pm 0.01 $	&	$ 609.9 \pm 3.4 $	&	$ 17.0 \pm 8.1 $	&	$ 0.3 \pm 0.2 $	\\
	&	$^{13}$CO(1-0)	&	$<0.03$	&	 \nodata	&	 \nodata	&	 \nodata	\\
	&	C$^{18}$O(2-1)	&	$<0.01$	&	 \nodata	&	 \nodata	&	 \nodata	\\
	&	C$^{18}$O(1-0)	&	$<0.04$	&	 \nodata	&	 \nodata	&	 \nodata	\\
	&	HCO$^{+}$(1-0)	&	$<0.07$	&	 \nodata	&	 \nodata	&	 \nodata	\\
	&	HCN(1-0)			&	$<0.06$	&	 \nodata	&	 \nodata	&	 \nodata	\\
	&	CN(1-0)			&	$<0.06$	&	 \nodata	&	 \nodata	&	 \nodata	\\
	&	HNC(1-0)			&	$<0.06$	&	 \nodata	&	 \nodata	&	 \nodata	\\
	&	CCH(1-0)			&	$<0.07$	&	 \nodata	&	 \nodata	&	 \nodata	\\

D-4	&	$^{12}$CO(2-1)	&	$ 0.13 \pm 0.05 $	&	$ 704.8 \pm 7.9 $	&	$ 44.0 \pm 18.0 $	&	$ 6.1 \pm 3.3 $	\\
	&	$^{13}$CO(2-1)	&	$<0.01$	&	 \nodata	&	 \nodata	&	 \nodata	\\
	&	$^{13}$CO(1-0)	&	$<0.03$	&	 \nodata	&	 \nodata	&	 \nodata	\\
	&	C$^{18}$O(2-1)	&	$<0.01$	&	 \nodata	&	 \nodata	&	 \nodata	\\
	&	C$^{18}$O(1-0)	&	$<0.04$	&	 \nodata	&	 \nodata	&	 \nodata	\\
	&	HCO$^{+}$(1-0)	&	$<0.07$	&	 \nodata	&	 \nodata	&	 \nodata	\\
	&	HCN(1-0)			&	$<0.06$	&	 \nodata	&	 \nodata	&	 \nodata	\\
	&	CN(1-0)			&	$<0.06$	&	 \nodata	&	 \nodata	&	 \nodata	\\
	&	HNC(1-0)			&	$<0.06$	&	 \nodata	&	 \nodata	&	 \nodata	\\
	&	CCH(1-0)			&	$<0.07$	&	 \nodata	&	 \nodata	&	 \nodata	\\

{\bf E-1	}&	$^{12}$CO(2-1)	&	$ 1.05 \pm 0.03 $	&	$ 374.5 \pm 1.0 $	&	$ 61.2 \pm 2.3 $	&	$ 68.3 \pm 3.3 $	\\
(13:25:27.918)	&	$^{13}$CO(2-1)	&	$ 0.07 \pm 0.01 $	&	$ 376.1 \pm 2.0 $	&	$ 49.3 \pm 4.7 $	&	$ 3.5 \pm 0.4 $	\\
(-43.01.15.132)	&	$^{13}$CO(1-0)	&	$<0.04$	&	 \nodata	&	 \nodata	&	 \nodata	\\
	&	C$^{18}$O(2-1)	&	$<0.01$	&	 \nodata	&	 \nodata	&	 \nodata	\\
	&	C$^{18}$O(1-0)	&	$<0.04$	&	 \nodata	&	 \nodata	&	 \nodata	\\
	&	HCO$^{+}$(1-0)	&	$ 0.42 \pm 0.02 $	&	$ 374.2 \pm 1.4 $	&	$ 52.8 \pm 3.4 $	&	$ 23.6 \pm 2.0 $	\\
	&	HCN(1-0)			&	$ 0.19 \pm 0.02 $	&	$ 372.8 \pm 1.2 $	&	$ 46.2 \pm 6.3 $	&	$ 9.5 \pm 1.7 $	\\
	&	CN(1-0)			&	$ 0.09 \pm 0.02 $	&	$ 371.2 \pm 1.8 $	&	$ 19.3 \pm 4.4 $	&	$ 1.8 \pm 0.5 $	\\
	&	HNC(1-0)			&	$ 0.05 \pm 0.01 $	&	$ 359.0 \pm 11.0 $	&	$ 105.0 \pm 31.0 $	&	$ 6.2 \pm 2.2 $	\\
	&	CCH(1-0)			&	$ 0.07 \pm 0.02 $	&	$ 399.1 \pm 5.3 $	&	$ 44.0 \pm 15.0 $	&	$ 3.3 \pm 1.4 $	\\

E-2	&	$^{12}$CO(2-1)	&	$<0.23$	&	 \nodata	&	 \nodata	&	 \nodata	\\
	&	$^{13}$CO(2-1)	&	$ 0.11 \pm 0.04 $	&	$ 543.6 \pm 1.6 $	&	$ 9.1 \pm 3.8 $	&	$ 1.0 \pm 0.6 $	\\
	&	$^{13}$CO(1-0)	&	$<0.04$	&	 \nodata	&	 \nodata	&	 \nodata	\\
	&	C$^{18}$O(2-1)	&	$<0.01$	&	 \nodata	&	 \nodata	&	 \nodata	\\
	&	C$^{18}$O(1-0)	&	$<0.04$	&	 \nodata	&	 \nodata	&	 \nodata	\\
	&	HCO$^{+}$(1-0)	&	$ 0.09 \pm 0.03 $	&	$ 544.6 \pm 2.1 $	&	$ 13.6 \pm 4.9 $	&	$ 1.3 \pm 0.6 $	\\
	&	HCN(1-0)			&	$<0.06$	&	 \nodata	&	 \nodata	&	 \nodata	\\
	&	CN(1-0)			&	$<0.04$	&	 \nodata	&	 \nodata	&	 \nodata	\\
	&	HNC(1-0)			&	$<0.06$	&	 \nodata	&	 \nodata	&	 \nodata	\\
	&	CCH(1-0)			&	$<0.07$	&	 \nodata	&	 \nodata	&	 \nodata	\\

E-3	&	$^{12}$CO(2-1)	&	$ 0.37 \pm 0.07 $	&	$ 590.0 \pm 3.7 $	&	$ 39.8 \pm 8.7 $	&	$ 15.9 \pm 4.6 $	\\
	&	$^{13}$CO(2-1)	&	$ 0.03 \pm 0.01 $	&	$ 584.0 \pm 1.2 $	&	$ 8.2 \pm 2.9 $		&	$ 0.3 \pm 0.1 $	\\
	&	$^{13}$CO(1-0)	&	$<0.04$	&	 \nodata	&	 \nodata	&	 \nodata	\\
	&	C$^{18}$O(2-1)	&	$<0.01$	&	 \nodata	&	 \nodata	&	 \nodata	\\
	&	C$^{18}$O(1-0)	&	$<0.04$	&	 \nodata	&	 \nodata	&	 \nodata	\\
	&	HCO$^{+}$(1-0)	&	$<0.06$	&	 \nodata	&	 \nodata	&	 \nodata	\\
	&	HCN(1-0)			&	$<0.06$	&	 \nodata	&	 \nodata	&	 \nodata	\\
	&	CN(1-0)			&	$<0.04$	&	 \nodata	&	 \nodata	&	 \nodata	\\
	&	HNC(1-0)			&	$<0.06$	&	 \nodata	&	 \nodata	&	 \nodata	\\
	&	CCH(1-0)			&	$<0.07$	&	 \nodata	&	 \nodata	&	 \nodata	\\

F-1	&	$^{12}$CO(2-1)	&	$<0.15$	&	 \nodata	&	 \nodata	&	 \nodata	\\
(13:25:26.819)	&	$^{13}$CO(2-1)	&	$ 0.19 \pm 0.02 $	&	$ 545.6 \pm 0.3 $	&	$ 5.3 \pm 0.9 $	&	$ 1.1 \pm 0.2 $	\\
(-43.01.13.585)	&	$^{13}$CO(1-0)	&	$ 0.16 \pm 0.03 $	&	$ 549.2 \pm 0.3 $	&	$ 2.3 \pm 0.5 $	&	$ 0.4 \pm 0.1 $	\\
	&	C$^{18}$O(2-1)	&	$<0.01$	&	 \nodata	&	 \nodata	&	 \nodata	\\
	&	C$^{18}$O(1-0)	&	$<0.04$	&	 \nodata	&	 \nodata	&	 \nodata	\\
	&	HCO$^{+}$(1-0)	&	$<0.06$	&	 \nodata	&	 \nodata	&	 \nodata	\\
	&	HCN(1-0)			&	$<0.06$	&	 \nodata	&	 \nodata	&	 \nodata	\\
	&	CN(1-0)			&	$<0.04$	&	 \nodata	&	 \nodata	&	 \nodata	\\
	&	HNC(1-0)			&	$<0.06$	&	 \nodata	&	 \nodata	&	 \nodata	\\
	&	CCH(1-0)			&	$<0.07$	&	 \nodata	&	 \nodata	&	 \nodata	\\

F-2	&	$^{12}$CO(2-1)	&	$ 1.79 \pm 0.11 $	&	$ 581.0 \pm 0.6 $	&	$ 18.4 \pm 1.3 $	&	$ 35.1 \pm 3.3 $	\\
	&	$^{13}$CO(2-1)	&	$ 0.51 \pm 0.03 $	&	$ 581.2 \pm 0.4 $	&	$ 16.6 \pm 1.0 $	&	$ 8.9 \pm 0.7 $	\\
	&	$^{13}$CO(1-0)	&	$ 0.47 \pm 0.03 $	&	$ 584.9 \pm 0.4 $	&	$ 11.5 \pm 0.9 $	&	$ 5.7 \pm 0.6 $	\\
	&	C$^{18}$O(2-1)	&	$ 0.06 \pm 0.01 $	&	$ 582.4 \pm 1.4 $	&	$ 13.3 \pm 3.3 $	&	$ 0.8 \pm 0.3 $	\\
	&	C$^{18}$O(1-0)	&	$<0.04$	&	 \nodata	&	 \nodata	&	 \nodata	\\
	&	HCO$^{+}$(1-0)	&	$<0.06$	&	 \nodata	&	 \nodata	&	 \nodata	\\
	&	HCN(1-0)			&	$<0.06$	&	 \nodata	&	 \nodata	&	 \nodata	\\
	&	CN(1-0)			&	$<0.04$	&	 \nodata	&	 \nodata	&	 \nodata	\\
	&	HNC(1-0)			&	$<0.06$	&	 \nodata	&	 \nodata	&	 \nodata	\\
	&	CCH(1-0)			&	$<0.07$	&	 \nodata	&	 \nodata	&	 \nodata	\\

G-1	&	$^{12}$CO(2-1)	&	$ 0.25 \pm 0.12 $	&	$ 442.4 \pm 5.4 $	&	$ 23.0 \pm 13.0 $	&	$ 6.1 \pm 4.6 $	\\
(13:25:28.556)	&	$^{13}$CO(2-1)	&	$ 0.05 \pm 0.01 $	&	$ 469.3 \pm 8.2 $	&	$ 46.0 \pm 20.0 $	&	$ 2.2 \pm 1.1 $	\\
(-43.01.21.565)	&	$^{13}$CO(1-0)	&	$<0.04$	&	 \nodata	&	 \nodata	&	 \nodata	\\
	&	C$^{18}$O(2-1)	&	$<0.01$	&	 \nodata	&	 \nodata	&	 \nodata	\\
	&	C$^{18}$O(1-0)	&	$<0.04$	&	 \nodata	&	 \nodata	&	 \nodata	\\
	&	HCO$^{+}$(1-0)	&	$<0.06$	&	 \nodata	&	 \nodata	&	 \nodata	\\
	&	HCN(1-0)			&	$<0.06$	&	 \nodata	&	 \nodata	&	 \nodata	\\
	&	CN(1-0)			&	$<0.04$	&	 \nodata	&	 \nodata	&	 \nodata	\\
	&	HNC(1-0)			&	$<0.06$	&	 \nodata	&	 \nodata	&	 \nodata	\\
	&	CCH(1-0)			&	$<0.07$	&	 \nodata	&	 \nodata	&	 \nodata	\\

G-2	&	$^{12}$CO(2-1)	&	$ 2.16 \pm 0.12 $	&	$ 507.6 \pm 0.6 $	&	$ 23.6 \pm 1.5 $	&	$ 54.5 \pm 4.6 $	\\
	&	$^{13}$CO(2-1)	&	$ 1.22 \pm 0.04 $	&	$ 509.8 \pm 0.3 $	&	$ 18.9 \pm 0.7 $	&	$ 24.6 \pm 1.3 $	\\
	&	$^{13}$CO(1-0)	&	$ 0.52 \pm 0.04 $	&	$ 511.0 \pm 0.6 $	&	$ 18.2 \pm 1.4 $	&	$ 10.1 \pm 1.0 $	\\
	&	C$^{18}$O(2-1)	&	$ 0.09 \pm 0.02 $	&	$ 512.5 \pm 1.8 $	&	$ 19.2 \pm 4.3 $	&	$ 1.9 \pm 0.6 $	\\
	&	C$^{18}$O(1-0)	&	$<0.04$	&	 \nodata	&	 \nodata	&	 \nodata	\\
	&	HCO$^{+}$(1-0)	&	$ 0.28 \pm 0.05 $	&	$ 503.9 \pm 0.9 $	&	$ 10.4 \pm 2.1 $	&	$ 3.1 \pm 0.8 $	\\
	&	HCN(1-0)			&	$ 0.14 \pm 0.03 $	&	$ 510.7 \pm 1.7 $	&	$ 17.6 \pm 4.1 $	&	$ 2.6 \pm 0.8 $	\\
	&	CN(1-0)			&	$<0.04$	&	 \nodata	&	 \nodata	&	 \nodata	\\
	&	HNC(1-0)			&	$<0.06$	&	 \nodata	&	 \nodata	&	 \nodata	\\
	&	CCH(1-0)			&	$<0.07$	&	 \nodata	&	 \nodata	&	 \nodata	\\

H-1	&	$^{12}$CO(2-1)	&	$ 2.00 \pm 0.10 $	&	$ 524.2 \pm 0.6 $	&	$ 22.3 \pm 1.3 $	&	$ 47.6 \pm 3.6 $	\\
(13:25:28.556)	&	$^{13}$CO(2-1)	&	$ 1.23 \pm 0.09 $	&	$ 521.7 \pm 0.6 $	&	$ 14.9 \pm 1.4 $	&	$ 19.6 \pm 2.3 $	\\
(-43.01.20.343)	&	$^{13}$CO(1-0)	&	$ 1.02 \pm 0.05 $	&	$ 523.8 \pm 0.3 $	&	$ 11.3 \pm 0.6 $	&	$ 12.2 \pm 0.9 $	\\
	&	C$^{18}$O(2-1)	&	$ 0.35 \pm 0.03 $	&	$ 518.4 \pm 0.3 $	&	$ 6.3 \pm 0.5 	$	&	$ 2.3 \pm 0.3 $	\\
	&	C$^{18}$O(1-0)	&	$<0.04$	&	 \nodata	&	 \nodata	&	 \nodata	\\
	&	HCO$^{+}$(1-0)	&	$ 0.17 \pm 0.03 $	&	$ 520.7 \pm 1.4 $	&	$ 15.3 \pm 3.3 $	&	$ 2.8 \pm 0.8 $	\\
	&	HCN(1-0)			&	$ 0.12 \pm 0.04 $	&	$ 521.1 \pm 2.7 $	&	$ 18.4 \pm 6.5 $	&	$ 2.4 \pm 1.1 $	\\
	&	CN(1-0)			&	$<0.04$	&	 \nodata	&	 \nodata	&	 \nodata	\\
	&	HNC(1-0)			&	$<0.06$	&	 \nodata	&	 \nodata	&	 \nodata	\\
	&	CCH(1-0)			&	$<0.07$	&	 \nodata	&	 \nodata	&	 \nodata	\\

H-2	&	$^{12}$CO(2-1)	&	$<0.08$	&	 \nodata	&	 \nodata	&	 \nodata	\\
	&	$^{13}$CO(2-1)	&	$<0.01$	&	 \nodata	&	 \nodata	&	 \nodata	\\
	&	$^{13}$CO(1-0)	&	$ 0.09 \pm 0.03 $	&	$ 559.9 \pm 0.9 $	&	$ 6.8 \pm 2.1 $	&	$ 0.7 \pm 0.3 $	\\
	&	C$^{18}$O(2-1)	&	$ 0.04 \pm 0.01 $	&	$ 558.1 \pm 1.0 $	&	$ 9.9 \pm 2.4 $	&	$ 0.4 \pm 0.1 $	\\
	&	C$^{18}$O(1-0)	&	$<0.04$	&	 \nodata	&	 \nodata	&	 \nodata	\\
	&	HCO$^{+}$(1-0)	&	$<0.07$	&	 \nodata	&	 \nodata	&	 \nodata	\\
	&	HCN(1-0)			&	$<0.06$	&	 \nodata	&	 \nodata	&	 \nodata	\\
	&	CN(1-0)			&	$<0.04$	&	 \nodata	&	 \nodata	&	 \nodata	\\
	&	HNC(1-0)			&	$<0.06$	&	 \nodata	&	 \nodata	&	 \nodata	\\
	&	CCH(1-0)			&	$<0.07$	&	 \nodata	&	 \nodata	&	 \nodata	\\

\enddata
\label{tab:lw}
\tablecomments{ For non-detected features, a 1$\sigma$ upper limit is given.  Spectral features attributed to the CND are in bold. }
\end{deluxetable}

\begin{deluxetable}{lcccccccccc}
\centering
\rotate
\tabletypesize{\scriptsize}
\tablecaption{ Ratios of Selected Molecular Transitions }
\tablecolumns{11}
\tablewidth{0pt}
\tablehead{
\colhead{ID} & \colhead{$\frac{^{12}\mathrm{CO(2-1)}}{^{13}\mathrm{CO(2-1)}}$} & \colhead{$\frac{^{12}\mathrm{CO(2-1)}}{\mathrm{C}^{18}\mathrm{O(2-1)}}$} & 
\colhead{$\frac{^{13}\mathrm{CO(2-1)}}{^{13}\mathrm{CO(1-0)}}$} & \colhead{$\frac{\mathrm{C}^{18}\mathrm{O(2-1)}}{\mathrm{C}^{18}\mathrm{O(1-0)}}$} & 
\colhead{$\frac{^{13}\mathrm{CO(2-1)}}{\mathrm{C}^{18}\mathrm{O(2-1)}}$} & \colhead{$\frac{^{13}\mathrm{CO(1-0)}}{\mathrm{HCN(1-0)}}$} & \colhead{$\frac{\mathrm{HCN(1-0)}}{\mathrm{HCO^{+}(1-0)}}$} & 
\colhead{$\frac{\mathrm{HCN(1-0)}}{\mathrm{HNC(1-0)}}$} & \colhead{$\frac{\mathrm{HCN(1-0)}}{\mathrm{CN(1-0)}}$} & \colhead{$\frac{\mathrm{HCN(1-0)}}{\mathrm{CCH(1-0)}}$}
}

\startdata
A-1 & 	$1.5 \pm 0.3$	 &	 $6.4 \pm 1.6$	 & 	$1.4 \pm 0.2$ 	& 	$>4.2$ 	& 	$4.2 \pm 0.9$ 	 &	 $>7.2$ 	 &	 \nodata	 &	  \nodata	 &	  \nodata	 &	  \nodata	 \\
A-2 &  	$1.2 \pm 0.4$	 &	 $>14.8$ 	 & 	$1.5 \pm 0.4$ 	& 	 \nodata 	& 	$>12.8$  &	$> 3.1$  &	  \nodata	 &	  \nodata	 &	  \nodata	 &	  \nodata	 \\
A-3 &  	$>7.7$       &	 $>7.4$  & 	$<1.0$  	& 	 \nodata 	& 	 \nodata 	 &	 $>0.3$  &	 $<0.8$ 	 &	  \nodata	 &	  \nodata	 &	  \nodata	 \\
B-1 &  	$5.1 \pm 0.8$	 &	 $111.9 \pm 61.8$	 & 	$1.5 \pm 0.3$ 	& 	$>0.5$ 	& 	$21.8 \pm 12.1$ 	 &	 $3.5 \pm 1.4$	 &	 $0.6 \pm 0.3$	 &	 $>1.4$ 	 &	$>1.8$ 	 &	 $>1.3$ 	 \\
B-2 &  	$2.2 \pm 0.7$	 &	 $>25.1$ 	 & 	$>3.1$  	& 	 \nodata 	& 	$>11.7$  	 &	  \nodata	 &	  \nodata	 &	  \nodata	 &	  \nodata	 &	  \nodata	 \\
{\bf B-3} &  	$20.0 \pm 3.3$	 &	 $>144.1$ 	 & 	$>1.9$  	& 	 \nodata 	& 	$>7.2$  	 &	 $<0.2$ 	 &	 $0.5 \pm 0.1$	 &	 $1.7 \pm 0.4$	 &	 $4.3 \pm 1.3$	 &	 $2.6 \pm 0.9$	 \\
C-1 &  	$<4.5$ 	 &	  \nodata	 & 	$>0.7$  	& 	 \nodata 	& 	$>2.9$  	 &	  \nodata	 &	  \nodata	 &	  \nodata	 &	  \nodata	 &	  \nodata	 \\
C-2 &  	$1.1 \pm 0.5$	 &	 $>21.2$ 	 & 	$>5.1$  	& 	 \nodata 	& 	$>19.6$  	 &	  \nodata	 &	  \nodata	 &	  \nodata	 &	  \nodata	 &	  \nodata	 \\
{\bf C-3} & 	$24.2 \pm 4.7$	 &	 $>114.9$ 	 & 	$>1.2$  	& 	 \nodata 	& 	$>4.7$  	 &	 $<0.1$ 	 &	 $0.6 \pm 0.1$	 &	 $2.0 \pm 0.3$	 &	 $1.7 \pm 0.2$	 &	 $3.9 \pm 0.8$	 \\
{\bf D-1} & 	$23.5 \pm 4.4$	 &	 $>99.4$ 	 & 	$>1.1$  	& 	 \nodata 	& 	$>4.2$  	 &	 $<0.1$ 	 &	 $0.6 \pm 0.1$	 &	 $3.0 \pm 0.4$	 &	 $2.4 \pm 0.4$	 &	 $>7.0$ 	 \\
D-2 & 	$>42.9$ 	 &	 $>41.4$	  & 	 \nodata 	& 	 \nodata 	& 	 \nodata 	 &	  \nodata	 &	  \nodata	 &	  \nodata	 &	  \nodata	 &	  \nodata	 \\
D-3 & 	$<13.1$ 	 &	  \nodata	 & 	$>0.5$  	& 	 \nodata 	& 	$>2.0$  	 &	  \nodata	 &	  \nodata	 &	  \nodata	 &	  \nodata	 &	  \nodata	 \\
D-4 & 	$>15.9$ 	 &	 $>15.4$ 	 & 	 \nodata 	& 	 \nodata 	& 	 \nodata 	 &	  \nodata	 &	  \nodata	 &	  \nodata	 &	  \nodata	 &	  \nodata	 \\
{\bf E-1} & 	$16.0 \pm 2.6$	 &	 $>101.7$ 	 & 	$>1.9$ 	& 	 \nodata 	& 	$>6.4$  	 &	 $<0.2$ 	 &	 $0.5 \pm 0.1$	 &	 $3.5 \pm 0.9$	 &	 $2.2 \pm 0.5$	 &	 $2.7 \pm 0.7$	 \\
E-2 & 	$<2.2$ 	 &	  \nodata	          & 	$>3.0$  	& 	 \nodata 	& 	$>10.3$  	 &	  \nodata	 &	 $<0.7$ 	 &	  \nodata	 &	  \nodata	 &	  \nodata	 \\
E-3 & 	$11.4 \pm 4.4$	 &	 $>36.2$ 	 & 	$>0.9$  	& 	 \nodata 	& 	$>3.2$  	 &	  \nodata	 &	 \nodata	 &	  \nodata	 &	  \nodata	 &	  \nodata	 \\
F-1 & 	$<0.8$ 	 &	  \nodata	          & 	$1.2 \pm 0.3$ 	& 	 \nodata 	& 	$>19.4$  	 &	 $>2.9$ 	 &	  \nodata	 &	  \nodata	 &	  \nodata	 &	  \nodata	 \\
F-2 & 	$3.5 \pm 0.6$	 &	 $31.1 \pm 8.2$	 & 	$1.1 \pm 0.2$ 	& 	$>1.6$  	& 	$8.8 \pm 2.3$ 	 &	 $>8.4$ 	 &	  \nodata	 &	  \nodata	 &	  \nodata	 &	  \nodata	 \\
G-1 & 	$5.6 \pm 3.1$	 &	 $>25.8$ 	 & 	$>1.2$  	& 	 \nodata 	& 	$>4.6$  	 &	  \nodata	 &	  \nodata	 &	  \nodata	 &	  \nodata	 &	  \nodata	 \\
G-2 & 	$1.8 \pm 0.3$         & 	 $22.9 \pm 5.7$	 & 	$2.4 \pm 0.3$ 	& 	$>2.6$ 	& 	$12.9 \pm 3.2$ 	 &	 $3.7 \pm 0.8$	 &	 $0.5 \pm 0.1$	 &	 $>2.4$ 	 &	 $>3.2$ 	 &	 $>2.1$ 	 \\
H-1 & 	$1.6 \pm 0.3$	 &	 $5.7 \pm 1.0$	 & 	$1.2 \pm 0.2$ 	& 	$>9.8$ 	& 	$3.5 \pm 0.6$	 &	 $8.4 \pm 2.7$	 &	 $0.7 \pm 0.3$	 &	 $>1.9$ 	 &	 $>3.1$ 	 &	 $>1.8$ 	 \\
H-2 & 	 \nodata	          &	 $<2.1$ 	 & 	$<0.1$  	& 	$>1.1$  	& 	$<0.3$  	 &	 $>1.5$ 	 &	  \nodata	 &	  \nodata	 &	  \nodata	 &	  \nodata	 \\

\enddata
\label{tab:ratio}
\tablecomments{T$_{pk}$ ratios of the emission features.   Limits are given for values where only a single transition was detected.  Spectral features attributed to the CND are in bold.  }

\end{deluxetable}

\begin{deluxetable}{lcccc}
\centering
\tabletypesize{\scriptsize}
\tablecaption{ Opacity Ratios for Selected Absoption Components}
\tablewidth{0pt}
\tablecolumns{5}
\tablehead{
\colhead{Ratio} & \colhead{Abs. Comp. 1} & \colhead{Abs. Comp. 2} & \colhead{Abs. Comp. 3} & \colhead{Broad Median}
 
}

\startdata
 & {\bf (539.7 km s$^{-1}$)} & {\bf (543.4 km s$^{-1}$)} & {\bf (549.7 km s$^{-1}$)} & {\bf (576-604 km s$^{-1}$)} \\
 & & & & \\
$\frac{^{12}\mathrm{CO(2-1)}}{^{13}\mathrm{CO(2-1)}}$			& $1.05^{b} \pm 0.01$	& $0.50^{b} \pm 0.01$	& $1.82 	\pm 0.02	$	& $\sim 12.5$			  \\
 & & & & \\
$\frac{^{13}\mathrm{CO(2-1)}}{^{13}\mathrm{CO(1-0)}}$			& $1.45 	\pm 0.01	$	& $2.08	\pm 0.02	$	& $1.01	\pm 0.01$		& $>2.5$				  \\
 & & & & \\
$\frac{^{12}\mathrm{CO(2-1)}}{\mathrm{C}^{18}\mathrm{O(2-1)}}$	& $46^{b} \pm 8	$	& $17^{b} \pm 1	$	& $53 	\pm 2$		& $>31$			  \\
 & & & & \\
$\frac{^{13}\mathrm{CO(2-1)}}{\mathrm{C}^{18}\mathrm{O(2-1)}}$	& $44 	\pm 8	$	& $34	 \pm 2	$	& $29	\pm 1$		& $>2.5$				  \\
 & & & & \\
$\frac{^{12}\mathrm{CO(2-1)}}{\mathrm{HCO^{+}(1-0)}}$			& $0.11^{b} \pm 0.01$	& $0.14^{b} \pm 0.01$	& $0.22	\pm 0.01$		& $0.27 \pm 0.01$				  \\
 & & & & \\	
$\frac{^{13}\mathrm{CO(1-0)}}{\mathrm{HCN(1-0)}}$			& $0.22 	\pm 0.01	$	& $0.12	\pm 0.01	$	& $0.50	\pm 0.01$		& $<0.04$				  \\
 & & & & \\
$\frac{\mathrm{HCN(1-0)}}{\mathrm{HCO^{+}(1-0)}}$ 			& $0.34 	\pm 0.01	$	& $1.17^{b} \pm 0.01$	& $ 0.24^{b} 	\pm 0.01$	& $ 0.21^{b} 	\pm 0.01$	  \\
 & & & & \\
$\frac{\mathrm{HCN(1-0)}}{\mathrm{HNC(1-0)}}$ 				& $ 2.89  \pm 0.01	$	& $ 3.77^{b} \pm 0.04$	& $ 3.00^{b} 	\pm 0.03$	& $ 3.3^{b} 	\pm 0.3$	  \\
 & & & & \\
$\frac{\mathrm{HCN(1-0)}}{\mathrm{CN(1-0)}}$ 				& $ 1.11  \pm 0.01	$	& $ 1.74^{b} \pm 0.02$	& $ 1.23^{b} \pm 0.01$	& $ 0.71^{b} 	\pm 0.03$	  \\
 & & & & \\
$\frac{\mathrm{HCN(1-0)}}{\mathrm{CCH(1-0)}}$				& $ 3.00  \pm 0.01	$	& $ 6.56^{b} \pm 0.07$	& $ 3.32^{b} \pm 0.03$	& $ 1.54^{b} 	\pm 0.08$	  \\
 & & & & \\
$\frac{\mathrm{CCH(1-0)}}{\mathrm{CN(1-0)}}$ 				& $ 0.37  \pm 0.01	$	& $ 0.26^{b} \pm 0.01$	& $ 0.37^{b} \pm 0.01$	& $ 0.46^{b} 	\pm 0.02$	  \\

\enddata
\label{tab:absratiodlux}
\tablecomments{Table of the absorption ratios taken to specific velocities.  Peak opacities were used for the narrow components and for the broad absorption component, a median was calculated between velocities 576 and 604 km s$^{-1}$.  For the transitions that contain hyperfine structure, the ratios only include the strongest component of the hyperfine.  $^{b}$ refers to a case where one or both of the absorption components have a hyperfine component blended with it. In the case of $^{12}$CO(2-1), the first two absorption components are blended together due to poor velocity resolution of the data. }

\end{deluxetable}

\end{document}